\documentclass[a4paper,fleqn,usenatbib]{mnras}

\usepackage{newtxtext,newtxmath}

\usepackage[T1]{fontenc}
\usepackage{ae,aecompl}

\usepackage{color}

\usepackage{graphicx}	
\usepackage{amsmath}	

\newcommand{\rmd}{\,\mathrm{d}}
\newcommand{\rmi}{\mathrm{i}}
\newcommand{\del}{\partial}
\newcommand{\bsxi}{\boldsymbol{\xi}}
\newcommand{\bscalL}{\boldsymbol{\mathcal{L}}}

\title[Magnetic field topology and obliquity]{Topology and obliquity of core magnetic fields in shaping seismic properties of slowly rotating evolved stars}

\author[S.~T.~Loi]{
Shyeh Tjing Loi\thanks{E-mail: stl36@cam.ac.uk}
\\
Department of Applied Mathematics and Theoretical Physics, University of Cambridge, Centre for Mathematical Sciences, Wilberforce Road, Cambridge CB3 0WA, UK
}

\date{Accepted XXX. Received YYY; in original form ZZZ}

\pubyear{2020}

\begin{document}
\label{firstpage}
\pagerange{\pageref{firstpage}--\pageref{lastpage}}
\maketitle

\begin{abstract}
  It is thought that magnetic fields must be present in the interiors of stars to resolve certain discrepancies between theory and observation (e.g.~angular momentum transport), but such fields are difficult to detect and characterise. Asteroseismology is a powerful technique for inferring the internal structures of stars by measuring their oscillation frequencies, and succeeds particularly with evolved stars, owing to their mixed modes, which are sensitive to the deep interior. The goal of this work is to present a phenomenological study of the combined effects of rotation and magnetism in evolved stars, where both are assumed weak enough that first-order perturbation theory applies, and we focus on the regime where Coriolis and Lorentz forces are comparable. Axisymmetric ``twisted-torus'' field configurations are used, which are confined to the core and allowed to be misaligned with respect to the rotation axis. Factors such as the field radius, topology and obliquity are examined. We observe that fields with finer-scale radial structure and/or smaller radial extent produce smaller contributions to the frequency shift. The interplay of rotation and magnetism is shown to be complex: we demonstrate that it is possible for nearly symmetric multiplets of apparently low multiplicity to arise even under a substantial field, which might falsely appear to rule out its presence. Our results suggest that proper modelling of rotation and magnetism, in a simultaneous fashion, may be required to draw robust conclusions about the existence/non-existence of a core magnetic field in any given object.
\end{abstract}




\section{Introduction}
Magnetism in stars can occur in regions where a dynamo is operating to actively generate the field, or they may be fossil fields, i.e.~accreted passively from the parent gas cloud \citep{Mestel2012}. Regions in a star where dynamo action can operate are those unstable to turbulent convection, which occurs in the cores of intermediate- to high-mass ($>1.2\,M_\odot$) main sequence stars, the envelopes of less massive main sequence stars, and the envelopes of red giants of all masses \citep{Maeder2008}. While it is possible to directly observe magnetic fields that penetrate the surface (e.g.~through spectropolarimetry), whether this be from dynamo action or a fossil field \citep{Donati2009, Wade2015, Shultz2018}, core magnetism is much harder to establish and probe. For this, indirect means have been sought in the form of numerical simulations \citep{Brun2005, Featherstone2009}, and asteroseismology \citep{Fuller2015, Stello2016}.

Asteroseismology is the technique of deducing a star's interior properties from analysis of its natural modes of oscillation. Applied first to the Sun (helioseismology), it has since been turned with great success to more distant stars, yielding numerous breakthroughs particularly in the context of evolved stars (subgiants and red giants). This owes to the existence of \textit{mixed modes} in such objects, which involve a coupling of fluid motions deep in the core (gravity waves/g-mode oscillations) to those at the surface (acoustic waves/p-mode oscillations). This has enabled core rotation rates \citep{Beck2011} and the presence of core helium burning \citep{Bedding2011} to be inferred. Signatures of strong (i.e.~dynamically significant) magnetic fields in the cores of evolved stars are also thought to manifest in the form of mode depression \citep{Mosser2012}, where resonant interactions between gravity and Alfv\'{e}n waves cause mode conversion and energy dissipation by phase mixing \citep{Lecoanet2017, Loi2018, Loi2020}. However, the exact origin of observed mode depression and its necessity to invoke magnetic fields remains controversial \citep{Garcia2014, Mosser2017a}.

Numerous open questions surround the existence and role of internal magnetic fields in stellar physics/evolution. It is speculated that they may be able to explain observed efficiencies of angular momentum transport \citep{Aerts2019}, which are much higher than non-magnetic processes can collectively account for, although the exact mechanism(s) are under debate \citep{Fuller2019, Hartogh2020, Takahashi2020}. Magnetic fields may also help resolve the problem of photospheric chemical abundances in evolved stars, where enhanced mixing within radiative zones is required \citep{Mathis2005, Busso2007}. It is therefore of interest to find ways of detecting and characterising magnetic fields in deep stellar interiors, for which mixed-mode asteroseismology of evolved stars offers great promise.

The mode depression phenomenon, which is exclusive to those red giant stars $>$1.2\,$M_\odot$ (previously able to host core dynamos when on the main sequence), occurs in only a fraction of these \citep[$\sim$50\% for those $>$1.6\,$M_\odot$;][]{Stello2016}. It is therefore to be wondered what explains the remaining stars: might they also have significant core fields but just below the critical threshold needed for mode depression, or is there a genuine dichotomy in the field strengths/properties? Since magnetic fields are not spherically symmetric, the Lorentz force lifts the degeneracy of modes with the same radial order $n$ and spherical harmonic degree $\ell$ but different azimuthal order $m$, producing frequency splitting and giving rise to a multiplet. Similarly, the Coriolis force (arising from rotation) also produces frequency splitting \citep{Beck2011, Mosser2015}. At low rotation rates and/or weak fields, this problem can be treated using first-order perturbation theory, which calculates the frequency correction to the associated unperturbed eigenmode assuming that the Coriolis and Lorentz forces are small compared to pressure and buoyancy. This approach has been used to treat solar p-modes \citep{Gough1990}, fundamental/low-order p-modes in Cepheid variables \citep{Shibahashi2000}, and r-modes in degenerate stars \citep{Morsink2002} under the influence of rotation and magnetism. For g-modes, the effects of a magnetic field were investigated by \citet{Hasan2005} for slowly pulsating B-stars, and \citet{Rashba2007} in the case of the Sun.

At larger rotation rates/field strengths, e.g.~where the rotation/Alfv\'{e}n frequency is a substantial fraction of the mode frequency, non-perturbative approaches must be used. For example, there is a body of recent work on rapidly rotating stars where the Coriolis force is included via the traditional approximation \citep{Buysschaert2018, Prat2019, Prat2020, VanBeeck2020}, and first-order perturbation theory is used for the Lorentz force only. This is relevant for massive main sequence stars, for which magnetic braking does not efficiently operate to slow down their rotation. In contrast, evolved stars rotate much more slowly due to the huge expansion of their envelopes upon leaving the main sequence, and so rotational effects may be treated perturbatively. As mentioned above, the existence/properties of magnetic fields in red giant cores is still largely speculative. While a sizeable fraction may have strong fields that explain their mode depression, detailed predictions of how weaker magnetism might impact the seismic properties of evolved stars have been lacking. In particular, there are few studies examining the simultaneous consequences of rotation and magnetism in these objects, although several works have attempted to quantify the effects of magnetism by itself \citep{Cantiello2016, Gomes2020, Loi2020a}. The recent works of \citet{Mathis2021} and \citet{Bugnet2021}, which have included the effects of aligned rotation in addition to a large-scale axisymmetric field, have begun to fill these gaps. However, all the above works have assumed field configurations that are simple and large scale. In contrast, the current work aims to treat the unexplored regime of more complex radial field structure, that may also be misaligned with respect to rotation (i.e.~non-axisymmetric).

It is the goal of this work to perform a phenomenological study into the combined effects of rotation and magnetism on mixed modes in evolved stars, where both Coriolis and Lorentz forces are assumed to be weak enough that first-order perturbation theory applies. The focus lies on low-degree mixed modes of short radial wavelength, and the regime where the two forces are comparable in strength. Axisymmetric ``twisted torus'' configurations for the magnetic field will be used, and allowed to be inclined with respect to the rotation axis. This is motivated by the knowledge that rotation and magnetic axes often do not coincide, as seen both in stars with fossil fields \citep{Henrichs2013, Braithwaite2017} and in objects with active dynamos including the Sun \citep{Gosling2007} and the Earth \citep{Merrill2010}. Rotation will be treated as uniform for simplicity; note that in the core region this would be physically justified as magnetic fields tend to enforce solid-body rotation \citep{Spruit1999}. Besides obliquity of the field, the impact of topology will also be examined, by considering twisted torus configurations of more complex radial structure. Note that no good knowledge exists about the most likely field topologies that may be found in red giant cores. Speculatively, given that the most well-observed dynamos exhibit periodic reversals on relatively short timescales \citep{Jacobs1994, Ossendrijver2003}, and simulations of stellar core dynamos show similar behaviour \citep{Brun2005}, one might imagine that as the convective core recedes over the main sequence it might leave behind magnetised shells of alternating sign. Whether or how such small-scale complex structure (if it can survive to later stages) might influence frequency splittings has not previously been studied. It is to be noted that configurations with larger spatial scales are lower in energy and expected to be more stable \citep{Broderick2008, Duez2010a}. However, we consider here the possibility that some dynamos may preferentially create fields of smaller-scale structure, which may or may not eventually collapse to a lower energy state.

This paper is structured as follows. In Section \ref{sec:models}, we introduce the stellar models and magnetic field configurations. In Section \ref{sec:methods}, we describe how the basic (unperturbed) eigenmodes were computed, and review relevant aspects of first-order perturbation theory. Results are presented in Section \ref{sec:results} and discussed further in Section \ref{sec:discuss}, which covers implications for asteroseismic inference and limitations of the framework. Finally, we conclude in Section \ref{sec:summary}.

\section{Models}\label{sec:models}
Four stellar models were examined in this study, all of mass $M_* = 2\,M_\odot$. Two are polytropes of differing index, and two are realistic evolved stellar models generated by the publicly available stellar evolutionary code `Modules for Experiments in Stellar Astrophysics' (\textsc{mesa}, version r11701) \citep{Paxton2011}. Their parameters are summarised in Table \ref{tab:models}, and described in further detail below. The structure of these models was computed neglecting rotation and magnetism, as it is assumed that these effects are too small to produce significant deformations away from sphericity, or otherwise influence the hydrostatic background structure. The magnetic field models were calculated separately (see Section \ref{sec:Prendergast} for details), scaled to a desired strength and imposed on the core region of each model within a boundary of radius $R_\text{f}$.

The rotation profile is assumed to be uniform and described by a single scalar frequency $\Omega$, whose value is indicated in Table \ref{tab:models} for each of the models. For the polytropic models (A and B), $\Omega$ was chosen to be a small fixed fraction (0.2\%) of the dynamical frequency $\omega_\text{dyn} = \sqrt{GM_*/R_*^3}$, where $R_*$ is the stellar radius. For the \textsc{mesa} models (C and D), $\Omega$ was chosen to be consistent with characteristic rotational splittings of $\sim$100\,nHz observed for dipole modes in red giant stars of around $2\,M_\odot$ \citep[][fig.~6]{Mosser2012b}. This translates to $\Omega$ values of about 1\% and 3\% of the dynamical frequencies of Models C and D, respectively.

\subsection{Polytropes}
Solutions to the Lane-Emden equation
\begin{align}
  \frac{1}{\chi^2} \frac{\rmd}{\rmd \chi} \left( \chi^2 \frac{\rmd \tau}{\rmd \chi} \right) + \tau^\eta = 0 \:, \label{eq:LaneEmden}
\end{align}
which comes from substituting the polytropic relation $p \propto \rho^{1+1/\eta}$ into the equations of hydrostatic equilibrium, provide simple models for stars. Here $p$ is the gas pressure, $\rho$ is the mass density, $\eta$ is the polytropic index, $\chi$ is the radial coordinate, and $\tau$ is the polytropic temperature. These models are easy to generate and were considered in this study to see how much the results depended on the structure of the background.

\begin{figure*}
  \centering
  \includegraphics[width=\textwidth]{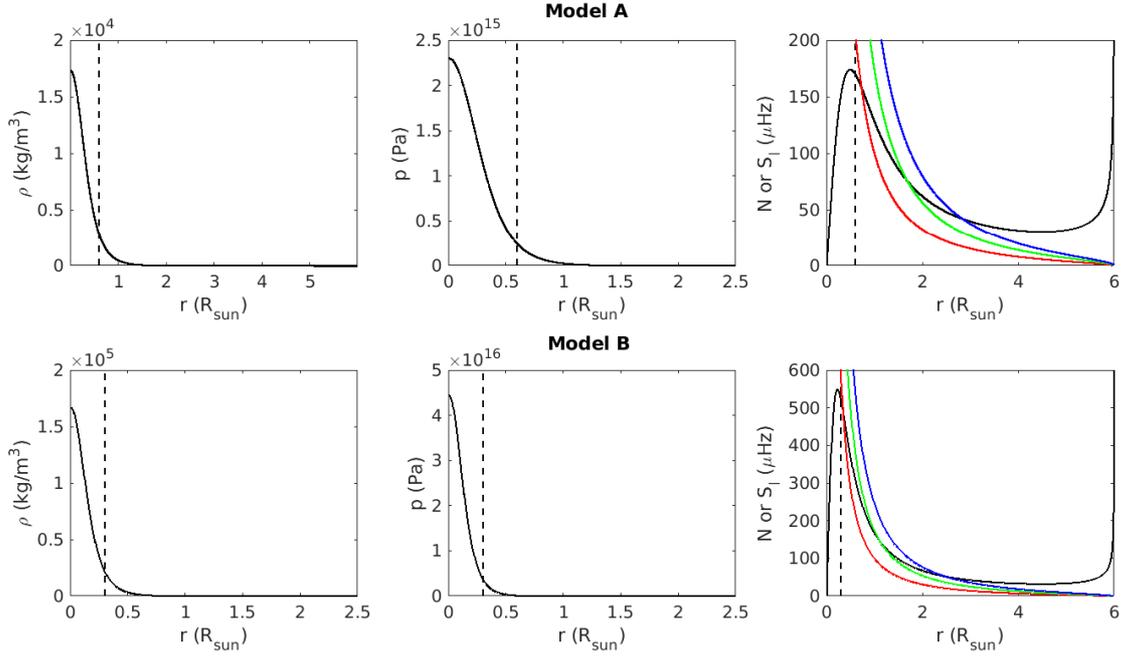}
  \caption{Background profiles showing the mass density $\rho$ (left), gas pressure $p$ (middle) and the Lamb and buoyancy frequencies $S_\ell$ and $N$ (right) for the two polytropic models (A and B), of index $\eta$ = 4.2 (top) and 4.6 (bottom). In the rightmost panels, $S_\ell$ is plotted in red, green and blue for $\ell$ = 1, 2 and 3, and $N$ is plotted in black. The dashed vertical lines indicate the default value of the field radius $R_\text{f}$ used in most of the calculations. Note that $S_\ell$ and $N$ are shown plotted up to the stellar surface, while the plots of $\rho$ and $p$ are zoomed in to the core region.}
  \label{fig:backgrounds_A_B}
\end{figure*}

Two polytropic models were generated, having the same mass $M_* = 2\,M_\odot$ and radius $R_* = 6\,R_\odot$, but different values of $\eta$. These are referred to as Models A and B, where A has $\eta = 4.2$ and B has $\eta = 4.6$. The value of $\eta$ controls the central condensation of a polytrope: Model B has a smaller, denser core and larger, more tenuous envelope than Model A. These are shown in Fig.~\ref{fig:backgrounds_A_B}, which shows the radial variation of $p$, $\rho$ and the Lamb and buoyancy frequencies, $S_\ell$ and $N$. A key difference between polytropes and realistic stellar models is that the former lack an evanescent zone between regions of p-mode and g-mode propagation, and are non-convective throughout. Thus all modes in polytropes are heavily mixed, with substantial amounts of both p- and g-like character. 

\subsection{Evolved stellar models}
\begin{figure*}
  \centering
  \includegraphics[width=\textwidth]{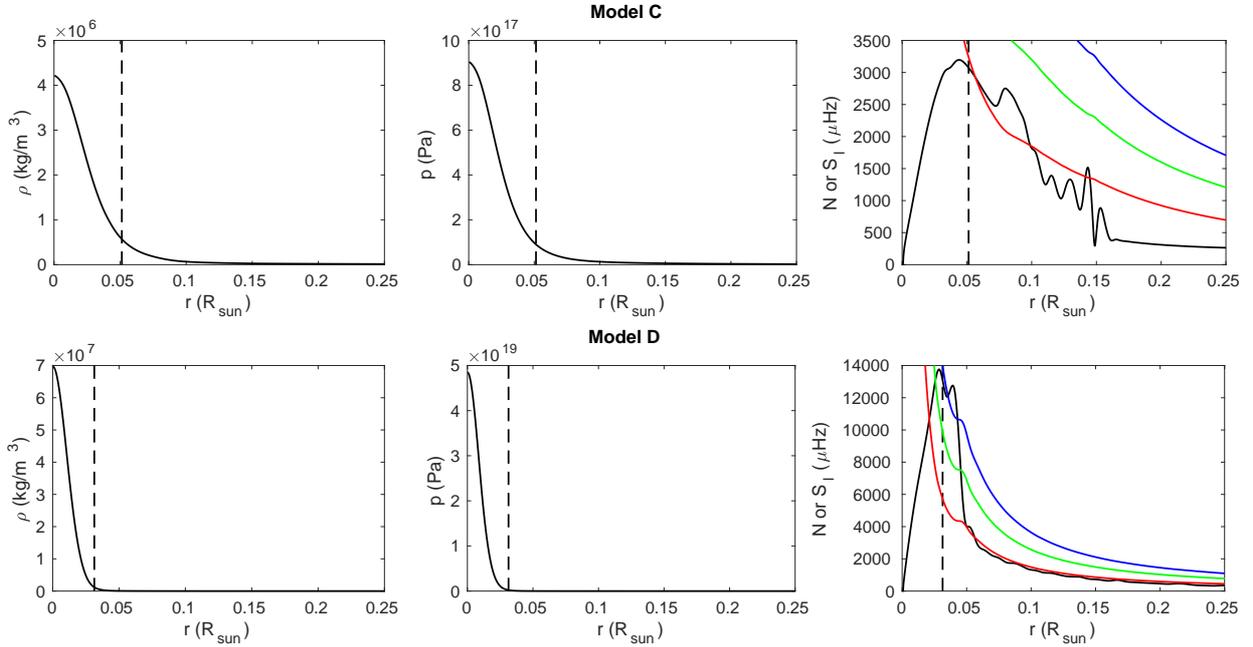}
  \caption{As in Fig.~\ref{fig:backgrounds_A_B}, but for the two \textsc{mesa} models (C and D). Note that plots of all quantities are zoomed in to the core region; the full stellar radii are 4.3 and 7.8\,$R_\odot$ for Models C and D, respectively.}
  \label{fig:backgrounds_C_D}
\end{figure*}

A sequence of 2\,$M_\odot$ stellar models (with metallicity $Z = 0.02$) along a single evolutionary track was generated using \textsc{mesa}, the inlist of which is given in Appendix \ref{sec:inlist}. Two of the output models were selected for further analysis, corresponding to a subgiant (Model C) and a red giant (Model D), at ages of 976\,Myr and 1.01 Gyr, respectively. It was noticed that under the default settings of \textsc{mesa} a significant level of jaggedness on the grid scale would be present. To remedy this, the mesh criteria were tightened to force \textsc{mesa} to use a finer grid, and further smoothing was applied to the outputs. The resultant profiles of $\rho$, $p$ and $N$ for Models C and D are shown in Fig.~\ref{fig:backgrounds_C_D}. 

In more detail, the smoothing procedure involved first replacing each point by that lying on the least-squares line of best fit through the 13 neighbouring points (cf.~boxcar smoothing), which were then linearly interpolated to a uniform grid having the same number of points. This was downsampled by a factor of 5, and then spline interpolated to a uniform grid having a chosen final number of points, which were $10^5$ and $10^6$ points for Models C and D, respectively. Note that the values listed under `\# grid points' in Table \ref{tab:models} are only half as large, as they refer to the size of the grid on which the eigenmodes were saved (this stems from the way the integration routine was implemented).

\subsection{Magnetic field}\label{sec:Prendergast}
Realistic modelling of magnetic fields in stars is a non-trivial task, owing to the need to satisfy various physical and stability constraints. Besides the well-known solenoidal condition $\nabla \cdot \mathbf{B} = 0$, where $\mathbf{B}$ is the magnetic field, all components of $\mathbf{B}$ need to be finite and continuous everywhere (to avoid infinite current sheets). If one desires to model a spatially confined field (e.g.~core of an evolved star) then such a field cannot in general be force-free \citep{Spruit2013}. Furthermore, any purely poloidal or purely toroidal configuration is unstable and so only mixed configurations are allowed \citep{Tayler1973, Flowers1977}. Most simple configurations, including uniform vertical fields, purely toroidal fields, and dipole fields, which have been widely used in many works, violate one or more of the above and are thus unlikely to be good descriptions of fields in reality. Instead, various analytical studies and numerical simulations point strongly towards so-called ``twisted-torus'' configurations \citep{Prendergast1956, Braithwaite2006, Yoshida2006, Duez2010a, Duez2010}, which are axisymmetric and dipole-like in angular appearance (in the sense that they have two poles) but differ in several important ways from actual dipoles, namely that they have no central singularity, vanish smoothly at a finite radius, and possess a stabilising toroidal component of comparable magnitude to the poloidal component. 

Such configurations, despite being moderately complex in appearance, are surprisingly easy to construct analytically. This was first achieved by \citet{Prendergast1956} for incompressible fluids and later generalised to the compressible case by \citet{Duez2010a} and \citet{Duez2010}. The derivation will not be covered here; the reader is encouraged to refer to the above works for details, also see \citet{Loi2020} for a summary. Although derived in the non-rotating case, numerical work by \citet{Duez2011} finds similar configurations in the presence of rotation, and these to be preferentially oblique with respect to the rotation axis. Such ``twisted-torus'' field configurations will be referred to here as \textit{Prendergast solutions}.

Each Prendergast solution is completely described by two quantities: a scalar function of radius, $\Psi(r)$, known as the radial flux function, and a parameter $\lambda$. These yield the magnetic field components according to
\begin{align}
  \mathbf{B} &= (B_r, B_{\theta'}, B_{\phi'}) \\
  &= \left( \frac{2\Psi}{r^2} \cos \theta', -\frac{1}{r} \frac{\rmd \Psi}{\rmd r} \sin \theta', -\frac{\lambda\Psi}{r} \sin \theta' \right) \:, \label{eq:Bfield}
\end{align}
where $(r, \theta', \phi')$ are spherical polar coordinates defined with respect to the magnetic axis. Primes are used to distinguish the angular coordinates from those defined with respect to the rotation axis (i.e.~$\theta$ and $\phi$), which is allowed to be different from the magnetic axis. The parameter $\lambda$ must be a root of
\begin{align}
  \int_0^{R_\text{f}} \rho \xi^3 j_1(\lambda \xi) \rmd \xi = 0 \:, \label{eq:lambda}
\end{align}
where $j_1$ is a spherical Bessel function of the first kind. Roots of (\ref{eq:lambda}) will exist if $\rho$ does not decrease too rapidly compared to $R_\text{f}$, and in general there can be multiple roots due to the oscillatory nature of $j_1$. Roughly speaking, $\lambda$ gives the inverse length scale of $\Psi$, and so larger $\lambda$ roots correspond to configurations with progressively finer radial structure. Once $\lambda$ is chosen, $\Psi$ is constructed as
\begin{align}
  \Psi(r) &\propto \frac{\lambda r}{j_1(\lambda R_\text{f})} \left[ f_\lambda(r, R_\text{f}) \int_0^r \rho \xi^3 j_1(\lambda \xi) \rmd \xi \right. \nonumber \\
    & \quad \left. + j_1(\lambda r) \int_r^{R_\text{f}} \rho \xi^3 f_\lambda(\xi, R_\text{f}) \rmd \xi \right] \:, \label{eq:Psi}
\end{align}
where $f_\lambda(r_1, r_2) \equiv j_1(\lambda r_2) y_1(\lambda r_1) - j_1(\lambda r_1) y_1(\lambda r_2)$, and $y_1$ is a spherical Bessel function of the second kind. The final desired field strength determines the scaling of $\Psi$.

\begin{figure*}
  \centering
  \includegraphics[width=\textwidth]{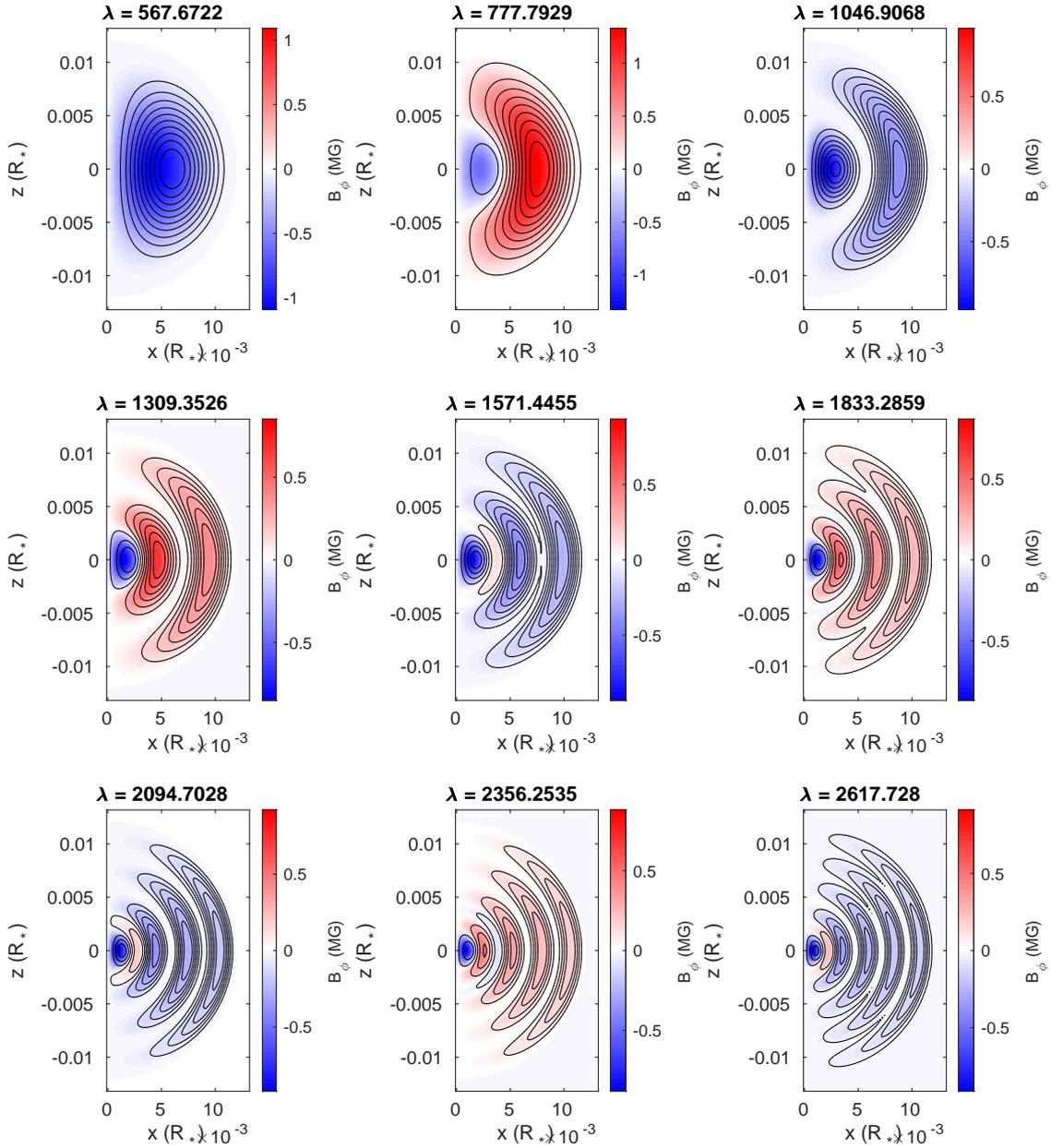}
  \caption{Prendergast solutions for the first nine $\lambda$ roots of Model C, shown in a meridional half-plane (note that the configurations are all axisymmetric). The vertical axis coincides with the axis of symmetry of the field, while the horizontal axis plots the cylindrical radius. Contours show poloidal projections of the field lines, while underlying colour indicates the strength of the toroidal component. Units on the colour bar are in MG. For the corresponding distributions of the total field strength, and similar plots for the other stellar models, see Supplementary Figs S1--S7.}
  \label{fig:Prendergast_C}
\end{figure*}

Prendergast solutions for the first nine $\lambda$ roots of Model C are shown in Fig.~\ref{fig:Prendergast_C}. While the lowest-order configuration may be described as a twisted torus, higher-order configurations take the form of multiple nested tori. Note how the north-south projection of the field lines flips sign between radial shells, which one might imagine could mimic the end result of a periodically reversing dynamo. Similar plots for the other stellar models, and corresponding distributions of the total field strength, are included as Supplementary Figs S1--S7. For the most part, the field strength is maximal at the centre, with additional smaller local maxima at larger radii. Along the boundary $r = R_\text{f}$, all field components smoothly match to the zero solution.

The default values of $R_\text{f}$ for each of the models are listed in Table \ref{tab:models}, and marked in Figs~\ref{fig:backgrounds_A_B} and \ref{fig:backgrounds_C_D}. In the case of the \textsc{mesa} models, they were chosen to lie within the boundary of the former convective core, which was identified through inspection of the H and He composition profiles. The overall scaling of the field strength was controlled through the central Alfv\'{e}n speed, the default values of which are also listed in Table \ref{tab:models}. These were chosen such that the magnetic contribution to the frequency splitting would be about the same order as that due to rotation. For comparison, the critical field strength \citep{Fuller2015, Loi2020} is also listed in the table, which marks the transition to dynamically significant field strengths where perturbation theory would be invalid.

\begin{table}
  \centering
  \caption{Summary of parameters for the four stellar models, including their mass $M_*$, radius $R_*$, dynamical frequency $\omega_\text{dyn} = \sqrt{GM_*/R_*^3}$, dynamical speed $v_\text{dyn} = \sqrt{GM_*/R_*}$, grid size, range of oscillation frequencies $\omega$ considered, number of modes found, rotation frequency $\Omega$, radial extent of the magnetic field $R_\text{f}$, central Alfv\'{e}n speed $v_\text{A,cen}$, corresponding value of the central field strength $B_\text{cen}$, and an estimate of the critical strength $B_\text{crit}$ for comparison. Note that values of $R_\text{f}$ and $v_\text{A,cen}$ shown are defaults. In some parts of this paper differing values are used, but where this occurs it will be explicitly stated. Otherwise, the default values should be assumed.} 
  \label{tab:models}
  \begin{tabular}{lcccc}
    \hline
    Model & A & B & C & D \\ \hline
    Type & Polytrope & Polytrope & \textsc{mesa} & \textsc{mesa} \\
    & ($\eta = 4.2$) & ($\eta = 4.6$) & (subgiant) & (red giant) \\
    $M_*/M_\odot$ & 2.00 & 2.00 & 2.00 & 2.00 \\
    $R_*/R_\odot$ & 6.00 & 6.00 & 4.28 & 7.84 \\
    $\omega_\text{dyn}/2\pi$ ($\upmu$Hz) & 9.64 & 9.64 & 16.0 & 6.45 \\
    $v_\text{dyn}$ (m\,s$^{-1}$) & $2.5 \times 10^5$ & $2.5 \times 10^5$ & $3.0 \times 10^5$ & $2.2 \times 10^5$ \\
    \# grid points & 9565 & 101033 & $5 \times 10^4$ & $5 \times 10^5$ \\
    $\omega/\omega_\text{dyn}$ & 0.5--1 & 0.5--1 & 8--12 & 8--12 \\ 
    \# modes & 23 ($\ell = 1$) & 72 ($\ell = 1$) & 13 ($\ell = 1$) & 71 ($\ell = 1$) \\
    & 40 ($\ell = 2$) & 124 ($\ell = 2$) & 19 ($\ell = 2$) & 121 ($\ell = 2$) \\
    & 56 ($\ell = 3$) & 175 ($\ell = 3$) & 26 ($\ell = 3$) & 170 ($\ell = 3$) \\
    $\Omega/\omega_\text{dyn}$ & 0.002 & 0.002 & 0.01 & 0.03 \\
    $R_\text{f}/R_*$ & 0.1 & 0.05 & 0.012 & 0.004 \\
    $v_\text{A,cen}/v_\text{dyn}$ & $10^{-4}$ & $2 \times 10^{-5}$ & $2 \times 10^{-4}$ & $2 \times 10^{-5}$ \\
    $B_\text{cen}$ (MG) & 0.04 & 0.02 & 1 & 0.4 \\
    $B_\text{crit}$ (MG) & 100 & 50 & 15 & 2 \\
    \hline
  \end{tabular}
\end{table}

\section{Methods}\label{sec:methods}
\subsection{Basic eigenmodes}
First, we need to obtain the set of basic eigenmodes that exist in the absence of rotation and magnetism. The time-dependent fluid displacement vector field for a normal mode of frequency $\omega$ can be written
\begin{align}
  \bsxi(r,\theta,\phi,t) = \left[ \xi_r(r,\theta,\phi) \hat{\mathbf{r}} + \xi_\theta(r,\theta,\phi) \hat{\boldsymbol{\theta}} + \xi_\phi(r,\theta,\phi) \hat{\boldsymbol{\phi}} \right] \exp(-\rmi \omega t) \:, \label{eq:xi_vec}
\end{align}
where
\begin{align}
  \xi_r(r,\theta,\phi) &= \sum_{\ell,m} R_{\ell,m}(r) Y_\ell^m(\theta,\phi) \:, \nonumber \\
  \xi_\theta(r,\theta,\phi) &= \sum_{\ell,m} H_{\ell,m}(r) \frac{\del}{\del \theta} Y_\ell^m(\theta,\phi) \:, \nonumber \\
  \xi_\phi(r,\theta,\phi) &= \sum_{\ell,m} \frac{H_{\ell,m}(r)}{\sin\theta} \frac{\del}{\del \phi} Y_\ell^m(\theta,\phi) \:. \label{eq:xi_compts}
\end{align}
Note that torsional motions have been neglected as the only restoring forces at this stage are pressure and buoyancy. Also, for spherically symmetric backgrounds $R_{\ell,m}$ and $H_{\ell,m}$ will not depend on $m$, so we shall write $R_{\ell,m} \to R_\ell$, $H_{\ell,m} \to H_\ell$.

Substituting into the equations of motion and linearising, and projecting out a single spherical harmonic, we get the equations of stellar oscillation
\begin{align}
  \frac{\rmd R_\ell}{\rmd r} &= \left( \frac{\ell(\ell+1)}{r} - \frac{\rho r}{\gamma p} \omega^2 \right) H_\ell - \left( \frac{2}{r} + \frac{1}{\gamma p} \frac{\rmd p}{\rmd r} \right) R_\ell \:, \label{eq:osc1} \\
  \frac{\rmd H_\ell}{\rmd r} &= \frac{1}{r} \left( 1 - \frac{N^2}{\omega^2} \right) R_\ell - \left( \rho N^2 \left( \frac{\rmd p}{\rmd r} \right)^{-1} + \frac{1}{r} \right) H_\ell \:. \label{eq:osc2}
\end{align}
Note that we have made the Cowling approximation (suitable for modes with small radial scales) thus reducing the system to second order, and neglected spatial variations of the adiabatic index $\gamma$.

The eigenmodes used for subsequent analysis in this work were obtained by directly solving Equations \ref{eq:osc1} and \ref{eq:osc2} under appropriate boundary conditions (regularity at $r = 0$, and vanishing Lagrangian pressure perturbation at the surface). They were solved via standard methods (shooting with a fourth-order Runge-Kutta scheme and matching at an intermediate radius) to obtain a set of unperturbed eigenmodes for spherical harmonic degrees $\ell$ = 1, 2 and 3 over chosen $\omega$ ranges. These were $[0.5,1]\,\omega_\text{dyn}$ for Models A and B, and $[8,12]\,\omega_\text{dyn}$ for Models C and D. The numbers of modes found within each interval are listed in Table \ref{tab:models}. The radial order $n$ of the various modes was computed using the \citet{Eckart1960} scheme.

\begin{figure*}
  \centering
  \includegraphics[width=\textwidth]{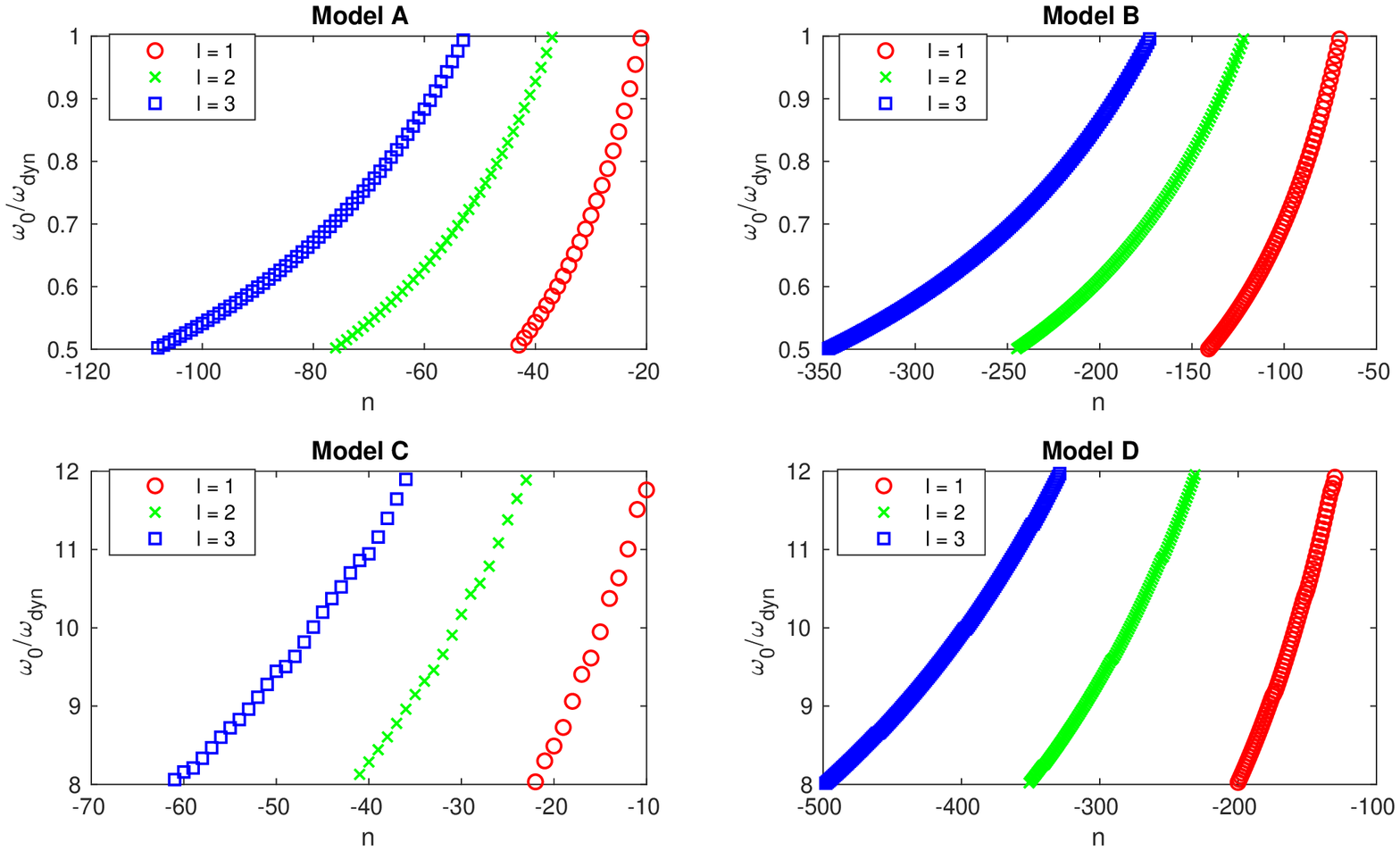}
  \caption{Frequencies $\omega_0$ (specified as a multiple of the dynamical frequency $\omega_\text{dyn}$) and radial orders $n$ of all modes with spherical harmonic degrees $\ell$ = 1, 2 and 3 obtained over the full search range. These are the modes in the spherically symmetric case (no rotation or magnetism).}
  \label{fig:modes}
\end{figure*}

\begin{figure}
  \centering
  \includegraphics[width=\columnwidth]{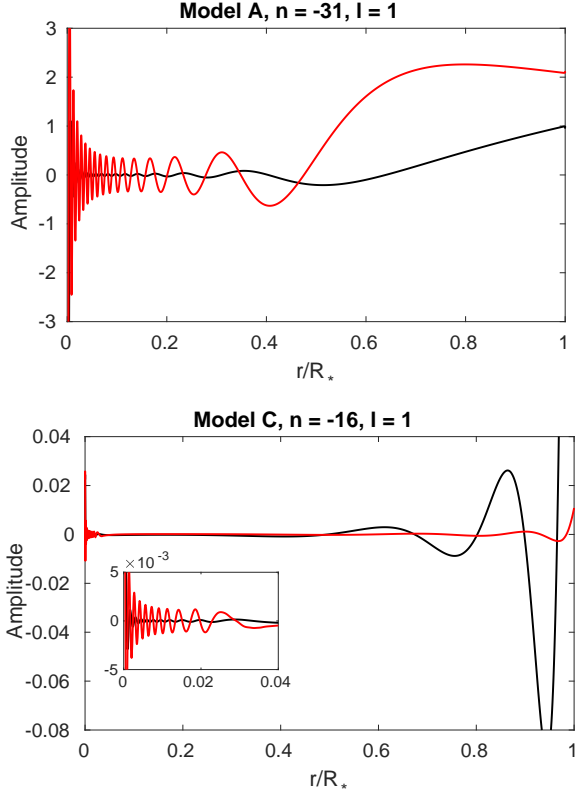}
  \caption{Selected eigenfunctions computed in the non-rotating, non-magnetic case (frequencies shown in Fig.~\ref{fig:modes}). Black and red correspond to the radial and horizontal components of the fluid displacement, $R_\ell$ and $H_\ell$. The associated model, $n$ and $\ell$ are indicated in the panel headings. The mixed character of the modes is evident from the simultaneously substantial core and surface displacements in both cases. In the lower panel, which is for a more centrally condensed model, the inset shows a zoom-in to the core region so that the g-mode oscillations can be better seen.}
  \label{fig:amplitude_fns}
\end{figure}

\begin{figure}
  \centering
  \includegraphics[width=\columnwidth]{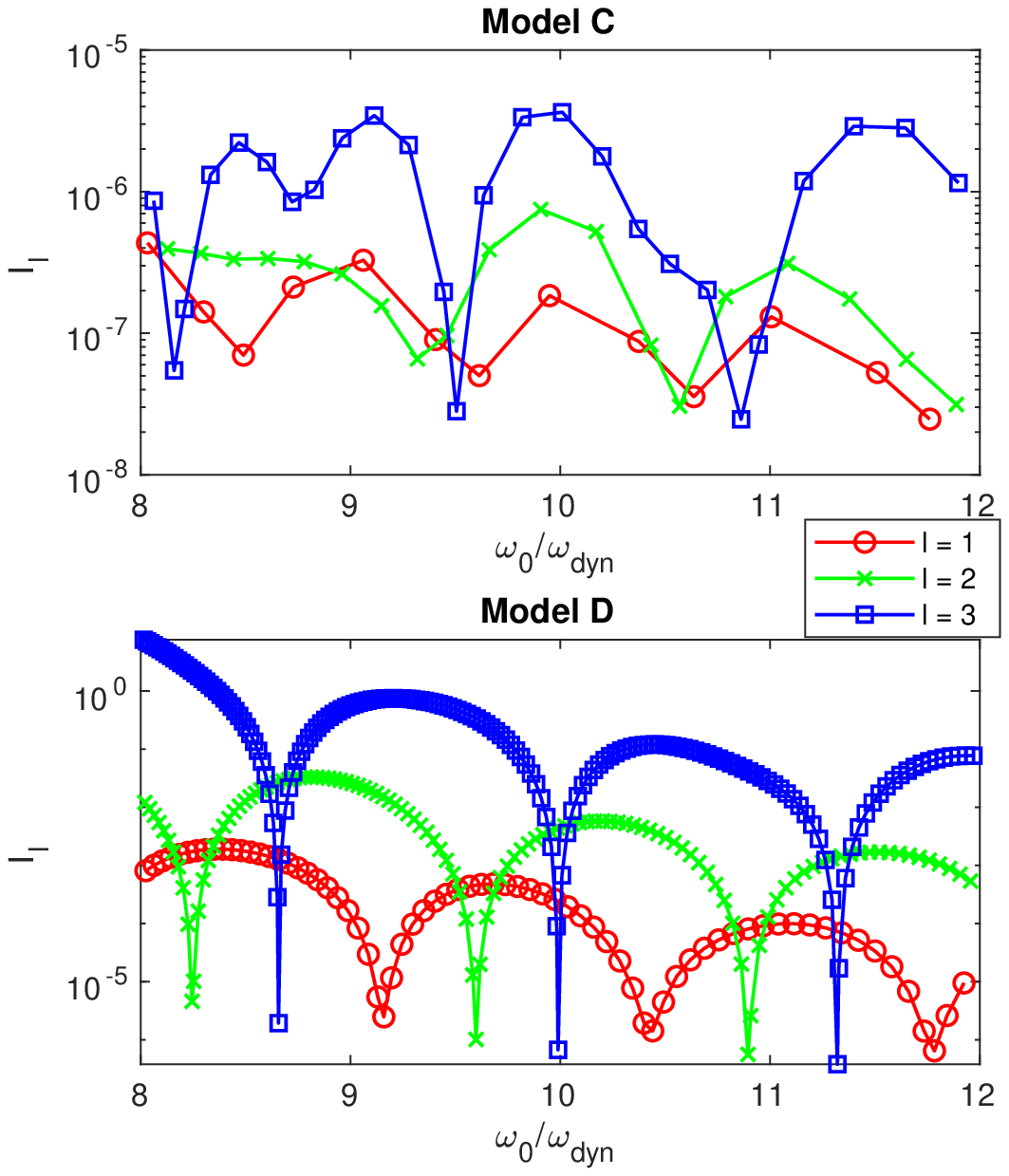}
  \caption{Mode inertias calculated according to (\ref{eq:inertia}), for the two \textsc{mesa} models. These show minima corresponding to where the mode becomes more p-dominated. In the more evolved model (bottom), the pure g-mode spectrum is much denser than the pure p-mode spectrum, leading to fewer minima in the mixed-mode spectrum per given number of modes, compared to Model C (top).}
  \label{fig:inertias}
\end{figure}

Figure \ref{fig:modes} plots the frequencies and radial orders of all modes. The vast majority of these are highly g-dominated, with large negative radial orders and frequencies that scale roughly as $1/|n|$. Examples of the associated eigenfunctions are shown in Fig.~\ref{fig:amplitude_fns}, where the finest scales of variation can be seen to occur in the deep interior. A useful measure of the p- or g-like character of a mode is the mode inertia
\begin{align}
  I_\ell = \frac{\int_0^{R_*} \rho r^2 \left[ R_\ell^2(r) + \ell(\ell+1) H_\ell^2(r) \right] \rmd r}{R_\ell^2(R_*) + \ell(\ell+1) H_\ell^2(R_*)} \:, \label{eq:inertia}
\end{align}
which measures the mass of fluid displaced and is larger for modes localised to the core where $\rho$ is larger, i.e.~g-dominated modes. Figure \ref{fig:inertias} shows the values of $I_\ell$ for all modes of Models C and D. The small number of modes with very low inertia correspond to the p-dominated modes, which occur near the frequencies of pure envelope p-modes. For Models A and B, their lack of an evanescent zone means that $I_\ell$ values are relatively constant over the considered frequency range; these are not shown.

\subsection{First-order perturbation theory}\label{sec:pertb}
To calculate frequency corrections to the basic eigenmodes induced by rotation and magnetism, first-order perturbation theory was used, which is a standard formalism e.g.~see \citet{Dziembowski1984, Unno1989}. It is valid in the limit of small perturbing forces, which is the regime focused on here. This section overviews relevant aspects of this theory, in particular its application to the problem of misaligned rotation and magnetism.

The presence of a non-spherically symmetric force (e.g.~Coriolis, Lorentz) lifts the degeneracy otherwise possessed by modes of the same $n$ and $\ell$ but different $m$. For a given $\ell$, there exist $2\ell + 1$ values of $m$ going from $-\ell$ to $+\ell$. To first order, pure rotation of frequency $\Omega$ perturbs the mode frequencies by a value proportional to $m\Omega$, giving a multiplet of $2\ell+1$ peaks. On the other hand, pure magnetism of characteristic field strength $B$ perturbs the frequencies by a value proportional to $B^2$ and independent of the sign of $m$, giving a multiplet of $\ell + 1$ peaks where modes of the same $|m|$ continue to experience a degeneracy. Note that this statement applies to axisymmetric fields, for which the quantum numbers $\ell$ and $m$ can be defined, and not for general non-axisymmetric fields where $\ell$ and $m$ would lose their meaning.

Under the combined influence of the two effects, the picture is more complicated. From the point of view of the star (i.e.~in the stationary/corotating frame), there can still be only $2\ell + 1$ frequencies since that is the number of originally degenerate modes. Under axisymmetric conditions (aligned rotation and magnetic axes) each of these is associated with a different $m$, and from the point of view of an observer (i.e.~in the inertial frame) there are thus $2\ell + 1$ peaks, where the frequencies are additionally Doppler shifted by a value of $m\Omega$. This is similar to the pure rotation case, except that due to the Lorentz force the multiplet may exhibit asymmetries; for a complete discussion and seismic diagnosis of these asymmetries, we refer the reader to the detailed discussion in \citet{Bugnet2021}. Under non-axisymmetric conditions (misaligned rotation and magnetic axes), there would still be $2\ell + 1$ modes in the stationary/corotating frame, but now each mode can no longer be associated with a single $m$ since the system has no axis of symmetry. Rather, each would be an admixture of the $2\ell + 1$ different values of $m$, giving rise to $(2\ell + 1)^2$ peaks in the inertial frame after Doppler shifting. Their amplitudes are directly related to the coefficients of expansion: for example, in the limit of mutual alignment (which is a special case of this more general framework), a single $m$ dominates strongly for each corotating-frame mode and thus only $2\ell + 1$ out of $(2\ell + 1)^2$ possible peaks have non-zero amplitudes.

The mathematical treatment is as follows. Let $m$ and $m'$ be defined as the azimuthal quantum numbers with respect to the rotation and magnetic axes, respectively, and let $\beta$ be the obliquity angle. Let $\omega$ and $\bar{\omega}$ be the frequencies in the inertial and corotating frames, respectively, and thus related by $\bar{\omega} = \omega + m\Omega$ (this definition means that $m < 0$ modes are prograde). The equation of motion can be written
\begin{align}
  \bar{\omega}^2 \bsxi = \bscalL_0 \bsxi + \bscalL_\text{rot} \bsxi + \bscalL_\text{mag} \bsxi \:, \label{eq:EoM}
\end{align}
where $\bscalL_\text{rot}$ and $\bscalL_\text{mag}$, which correspond to the Coriolis and Lorentz forces, are assumed to be much smaller than $\bscalL_0$, which represents the combination of pressure and buoyancy. Their functional forms in terms of the fluid displacement $\bsxi$ and background quantities are
\begin{align}
  \bscalL_0 \bsxi &= \frac{\nabla p}{\rho^2} \left( \rho \nabla \cdot \bsxi + \bsxi \cdot \nabla \rho \right) - \frac{1}{\rho} \nabla \left( \gamma p \nabla \cdot \bsxi + \bsxi \cdot \nabla p \right) \:, \label{eq:L0} \\
  \bscalL_\text{rot} \bsxi &= 2\rmi \bar{\omega} \boldsymbol{\Omega} \times \bsxi + \boldsymbol{\Omega} \times (\boldsymbol{\Omega} \times \bsxi) \:, \label{eq:Lrot} \\
  \bscalL_\text{mag} \bsxi &= \frac{1}{\rho} \mathbf{B} \times \left\{ \nabla \times \left[ (\mathbf{B} \cdot \nabla) \bsxi - \mathbf{B} (\nabla \cdot \bsxi) - (\bsxi \cdot \nabla) \mathbf{B} \right] \right\} \nonumber \\
  &\quad + \frac{1}{\rho} \left[ \nabla \times (\bsxi \times \mathbf{B}) \right] \times (\nabla \times \mathbf{B}) \nonumber \\
  &\quad - \frac{1}{\rho^2} \left( \rho \nabla \cdot \bsxi + \bsxi \cdot \nabla \rho \right) (\nabla \times \mathbf{B}) \times \mathbf{B} \:. \label{eq:Lmag}
\end{align}

The goal is to solve for the first-order frequency corrections $\bar{\omega}_1$, where $\bar{\omega} = \omega_0 + \bar{\omega}_1 + \cdots$. Here $\omega_0$ is the frequency associated with the unperturbed eigenmode $\bsxi_0$, where $\bsxi = \bsxi_0 + \bsxi_1 + \cdots$. Substituting these expansions into (\ref{eq:EoM}) and truncating at first order gives
\begin{align}
  \bscalL_1 \bsxi_0 + \bscalL_0 \bsxi_1 = \omega_0^2 \bsxi_1 + 2\omega_0 \bar{\omega}_1 \bsxi_0 \:, \label{eq:EoM_ptb1}
\end{align}
where $\bscalL_1 = \bscalL_\text{rot} + \bscalL_\text{mag}$. Now $\bscalL_0$ is self-adjoint, implying that $\{\bsxi_0\}$ form a complete orthogonal basis. Therefore, it is possible to express any $\bsxi_1 = \sum_k c_k \bsxi_0^{(k)}$, where $k$ is some enumeration of the set of $\{ \bsxi_0 \}$. Combining this with the fact that $\omega_0^2 \bsxi_0 = \bscalL_0 \bsxi_0$,
\begin{align}
  \bscalL_1 \bsxi_0^{(j)} + \left[ \sum_k c_k^{(j)} \left( \omega_0^{(k)2} - \omega_0^{(j)2} \right) \bsxi_0^{(k)} \right] = 2 \omega_0^{(j)} \bar{\omega}_1^{(j)} \bsxi_0^{(j)} \label{eq:EoM_ptb2}
\end{align}
for some eigenmode labelled $j$ (with chosen $n$ and $\ell$).

If $\bscalL_1$ were axisymmetric, it would suffice to consider $\bsxi_0^{(j)}$ as being made up of a single $m$. Then, we would be able to take the inner product of (\ref{eq:EoM_ptb2}) with $\bsxi_{0,m}^{(j)}$, which is defined to be (\ref{eq:xi_vec}) but restricted to a fixed $m$, and use orthogonality to get
\begin{align}
  \bar{\omega}_{1,m}^{(j)} = \frac{1}{2\omega_0^{(j)}} \frac{\left\langle \bscalL_1 \bsxi_{0,m}^{(j)} \:,\; \bsxi_{0,m}^{(j)} \right\rangle}{\left\langle \bsxi_0^{(j)} \:,\; \bsxi_0^{(j)} \right\rangle} \:. \label{eq:om1_aligned}
\end{align}
The inner product is defined as $\langle \bsxi, \boldsymbol{\eta} \rangle = \int \rho \bsxi^* \cdot \boldsymbol{\eta} \rmd V$, where the integral is over the volume of the star. However, if $\beta \neq 0$ then $\bscalL_1$ will not be axisymmetric, even if $\bscalL_\text{rot}$ and $\bscalL_\text{mag}$ individually are. In this case $\bscalL_1$ mixes the different $m$, and (\ref{eq:om1_aligned}) needs to be replaced by a $(2\ell+1) \times (2\ell+1)$ matrix equation. Let us write
\begin{align}
  \bsxi_0(r,\theta,\phi,t) = \sum_{m = -\ell}^{+\ell} a_m \bsxi_{0,m} = \bsxi'_0(r,\theta',\phi',t) = \sum_{m' = -\ell}^{+\ell} a'_{m'} \bsxi'_{0,m'} \:, \label{eq:trafo}
\end{align}
where $\bsxi'_{0,m'}$ has the same form as $\bsxi_{0,m}$ but with $(m, \theta, \phi) \to (m', \theta', \phi')$, and we have dropped the labels $j$ with the understanding that all quantities now pertain to the mode $j$.

To transform between coordinate frames $(r, \theta, \phi)$ and $(r, \theta', \phi')$ we invoke the Wigner $d$-matrix, denoted here by $\mathbf{D}$. Its entries are given by
\begin{align}
  d_{mm'}^{(\ell)}(\beta) &= \left[ (\ell+m)! (\ell-m)! (\ell+m')! (\ell-m')! \right]^{1/2} \nonumber \\
  &\times \sum_s \left[ \frac{(-1)^{m-m'+s} \left( \cos \frac{\beta}{2} \right)^{2\ell+m'-m-2s} \left( \sin \frac{\beta}{2} \right)^{m-m'+2s}}{(\ell+m'-s)! s! (m-m'+s)! (\ell-m-s)!} \right] \:, \label{eq:dmm}
\end{align}
where the summation is such that the factorials are non-negative. These satisfy
\begin{align}
  Y_\ell^m(\theta, \phi) &= \sum_{m' = -\ell}^{+\ell} d_{mm'}^{(\ell)}(\beta)\: Y_\ell^{m'}(\theta', \phi') \:,
\end{align}
which when substituted into (\ref{eq:xi_compts}) leads to
\begin{align}
  \bsxi_{0,m}(r,\theta,\phi,t) &= \sum_{m' = -\ell}^{+\ell} d_{mm'}^{(\ell)}(\beta)\: \bsxi'_{0,m'}(r,\theta',\phi',t) \:.
\end{align}
Hence (\ref{eq:trafo}) can be expressed as $\mathbf{a} = \mathbf{D} \mathbf{a}'$, where $\mathbf{a} = (a_m: m = -\ell, \cdots, +\ell)$ and $\mathbf{a}' = (a'_{m'}: m' = -\ell, \cdots, +\ell)$. We can then recast (\ref{eq:om1_aligned}) in the more general form of a matrix eigenvalue problem:
\begin{align}
  \bar{\omega}_1 \mathbf{a} = \boldsymbol{\mathcal{M}}_\text{rot} \mathbf{a} + \mathbf{D} \boldsymbol{\mathcal{M}}_\text{mag} \mathbf{D}^\top \mathbf{a} \:, \label{eq:om1_oblique}
\end{align}
where the Wigner $d$-matrix has the property that $\mathbf{D}^{-1} = \mathbf{D}^\top$, and
\begin{align}
  \boldsymbol{\mathcal{M}}_\text{rot} &= \frac{\text{diag}\left( \langle \bscalL_\text{rot} \bsxi_{0,m} \:,\; \bsxi_{0,m} \rangle \,:\, m = -\ell, \cdots, +\ell \right)}{2\omega_0 \langle \bsxi_0 \:,\; \bsxi_0 \rangle} \:, \label{eq:Mrot} \\
  \boldsymbol{\mathcal{M}}_\text{mag} &= \frac{\text{diag}\left( \langle \bscalL_\text{mag} \bsxi'_{0,m'} \:,\; \bsxi'_{0,m'} \rangle \,:\, m' = -\ell, \cdots, +\ell \right)}{2\omega_0 \langle \bsxi'_0 \:,\; \bsxi'_0 \rangle} \:. \label{eq:Mmag}
\end{align}
It is easy to see that this reduces to the aligned case when $\beta = 0 \implies \mathbf{D} = \mathbf{I}$, in which case the eigenvectors $\mathbf{a}$ are simply the columns of $\mathbf{I}$. The vectorial nature of $\mathbf{a}$ is not to be confused with a spatial coordinate vector; rather, it is a coordinate representation of the admixture of $m$, with $2\ell + 1$ components.

A word of caution is to be made about the self-adjointness of the Lorentz operator $\bscalL_\text{mag}$. In the fully self-consistent case where deformation is taken into account, the combined operator in (\ref{eq:EoM}) would be self-adjoint thus producing real frequencies, which is expected in ideal MHD. Note that in practice this deformation is small, and its dynamical consequences can largely be neglected \citep{Duez2010b}. As it stands, $\bscalL_0$ and $\bscalL_\text{rot}$ defined in (\ref{eq:L0}) and (\ref{eq:Lrot}) are self-adjoint, but $\bscalL_\text{mag}$ as defined in (\ref{eq:Lmag}) is not. The approximations used thus have the side effect of introducing imaginary components to the frequencies which are not physical. However, there exists a convenient remedy in the form of retaining only the first term on the RHS of (\ref{eq:Lmag}), which is the dominant term for modes of short wavelength and also happens to be self-adjoint. Hence we make the approximation
\begin{align}
  \bscalL_\text{mag} \bsxi \approx \frac{1}{\rho} \mathbf{B} \times \left\{ \nabla \times \left[ (\mathbf{B} \cdot \nabla) \bsxi - \mathbf{B} (\nabla \cdot \bsxi) - (\bsxi \cdot \nabla) \mathbf{B} \right] \right\} \:.
\end{align}
Also, since rotation frequencies are assumed to be small, we neglect the second (centrifugal) term in $\bscalL_\text{rot}$ in favour of the first (Coriolis):
\begin{align}
  \bscalL_\text{rot} \bsxi \approx 2\rmi \omega_0 \boldsymbol{\Omega} \times \bsxi
\end{align}
The $\bscalL_0$ term is left as is. With these simplifications, the elements of $\boldsymbol{\mathcal{M}}_\text{rot}$ and $\boldsymbol{\mathcal{M}}_\text{mag}$ are then
\begin{align}
  \langle \bscalL_\text{rot} \bsxi_{0,m} \:,\; \bsxi_{0,m} \rangle &= 4\omega_0 \int \rho \Omega \;\text{Im} \left[ \xi_\phi \left( \xi_\theta^* \cos \theta + \xi_r^* \sin \theta \right) \right] \rmd V \:, \label{eq:Mrot_el} \\
  \langle \bscalL_\text{mag} \bsxi'_{0,m'} \:,\; \bsxi'_{0,m'} \rangle &= \int \left[ \left( B_{\theta'} J_{\phi'}'^* - B_{\phi'} J_{\theta'}'^* \right) \xi'_r \right. \nonumber \\
    &\quad + \left( B_{\phi'} J_r'^* - B_r J_{\phi'}'^* \right) \xi'_{\theta'} \nonumber \\
    &\quad + \left. \left( B_r J_{\theta'}'^* - B_{\theta'} J_r'^* \right) \xi'_{\phi'} \right] \rmd V \:, \label{eq:Mmag_el}
\end{align}
where $\xi_r, \xi_\theta, \xi_\phi$ in (\ref{eq:Mrot_el}) refer to the components of $\bsxi_{0,m}$, and $\xi'_r, \xi'_{\theta'}, \xi'_{\phi'}$ in (\ref{eq:Mmag_el}) refer to the components of $\bsxi'_{0,m'}$. Here $\mathbf{J}' = (J_r', J_{\theta'}', J_{\phi'}')$ is the Eulerian perturbation to the current density. Expressions for the components of $\mathbf{J}'$ for the special case of the Prendergast solution can be found in Appendix \ref{sec:Jcompts}.

\section{Results}\label{sec:results}
\subsection{Comparison with pure rotation}\label{sec:results_rot}
\begin{figure}
  \centering
  \includegraphics[width=\columnwidth]{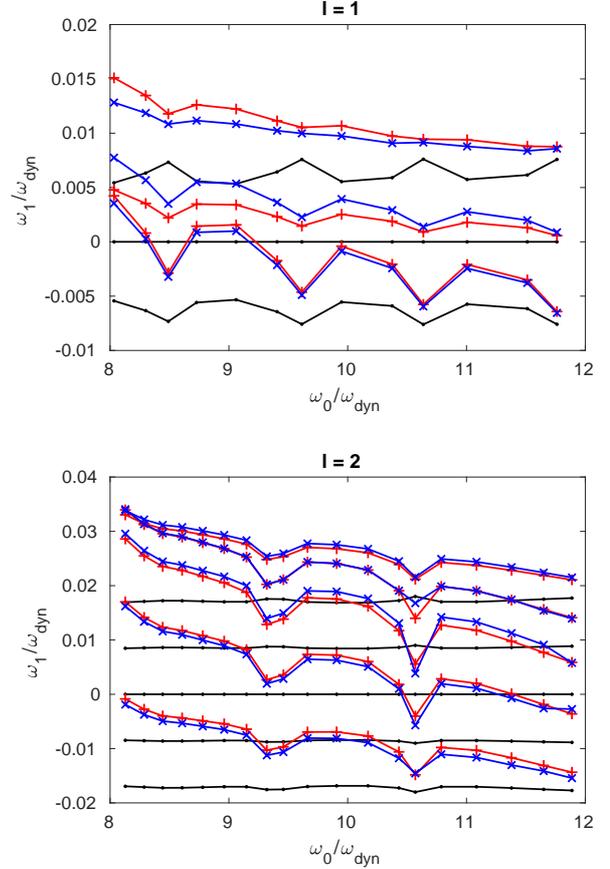}
  \caption{Inertial-frame frequency shifts versus unperturbed mode frequency (expressed as a multiple of the dynamical frequency) for Model C, where the different colours correspond to three cases: black dots are for zero field, red plusses are for non-zero field (of the default strength listed in Table \ref{tab:models}) with $\beta = 0$, and blue crosses are for the same field strength but with $\beta = \pi/4$. In all three cases the default rotation profile has been applied, which is $\Omega = 0.01\,\omega_\text{dyn}$ for this model. The $2\ell+1$ curves of the same colour in each panel correspond to the different modes in the multiplet; for the blue case the component with the largest $|a_m|$ has been selected for plotting.}
  \label{fig:comp_nofield_C_l12}
\end{figure}

\begin{figure}
  \centering
  \includegraphics[width=\columnwidth]{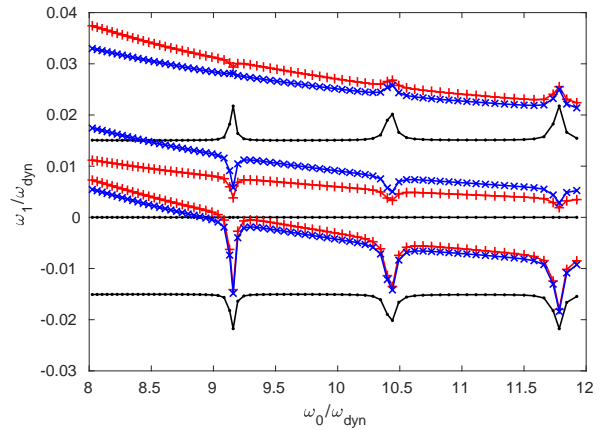}
  \caption{As for Fig.~\ref{fig:comp_nofield_C_l12}, but Model D and $\ell = 1$.}
  \label{fig:comp_nofield_D_l1}
\end{figure}

Figures \ref{fig:comp_nofield_C_l12} and \ref{fig:comp_nofield_D_l1} show the inertial-frame frequency shifts $\omega_1 = \bar{\omega}_1 - m\Omega$ for Models C and D as a function of $\omega_0$, for three different cases: pure rotation (black), rotation plus magnetic field where $\beta = 0$ (red), and rotation plus magnetic field where $\beta = \pi/4$ (blue). Comparing the red curve with the black curve, two aspects can be noted: (i) p-dominated multiplets are more symmetric than g-dominated ones, and (ii) the asymmetries are more pronounced at lower frequencies. The first point can be explained by the fact that the magnetic field is confined to the core where g-dominated mixed modes are localised; p-dominated mixed modes are thus more similar to modes in the case of pure rotation, for which $\omega_1$ is roughly constant with $\omega_0$. The second point can be understood through the frequency dependence of the elements of $\boldsymbol{\mathcal{M}}_\text{rot}$ and $\boldsymbol{\mathcal{M}}_\text{mag}$. While $\boldsymbol{\mathcal{M}}_\text{mag} \propto 1/\omega_0$, the factors of $\omega_0$ in $\boldsymbol{\mathcal{M}}_\text{rot}$ cancel out in the numerator and denominator, leaving no explicit $\omega_0$ dependence. The increased importance of magnetic effects at lower frequencies can be physically understood in terms of the increase of the Alfv\'{e}n frequency $\omega_A \propto k$ where $k$ is the wavenumber, thus bringing it closer to the mode frequency. Note that we are in the regime where $\omega_A \ll \omega_0$, and that $\omega_A \sim \omega_0$ corresponds to the strong-field (dynamically significant) regime in which perturbation theory breaks down. 

It is curious to note that for $\ell = 1$, in both Models C and D, the blue curves ($\beta = \pi/4$) exhibit more symmetric splittings than the red ($\beta = 0$). This appears to roughly hold across the whole frequency range. However, the $\ell = 2$ case does not show this behaviour, at least for the value of $\beta$ tested.

\subsection{Effect of obliquity}\label{sec:results_ob}
As mentioned in Section \ref{sec:pertb}, an important consequence of obliquity is to mix the different $m$ components, giving rise to $(2\ell + 1)^2$ frequencies in the inertial frame. However, only $2\ell + 1$ blue curves have been chosen for plotting in Figs~\ref{fig:comp_nofield_C_l12} and \ref{fig:comp_nofield_D_l1}; these correspond to the $m$ components with the largest $|a_m|$ values, i.e.~largest observed amplitudes. As the field strength and/or obliquity increase, it becomes less easy to identify the $m$ component with the largest $|a_m|$, as the various $|a_m|$ values become comparable.

\begin{figure}
  \centering
  \includegraphics[width=\columnwidth]{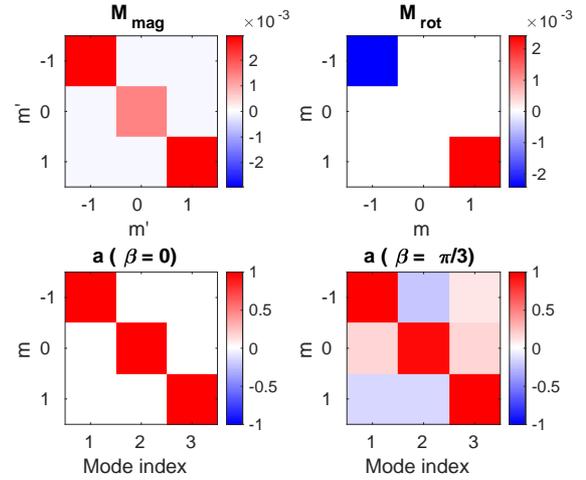}
  \caption{Top row: structure of the matrices $\mathcal{M}_\text{mag}$ and $\mathcal{M}_\text{rot}$, whose diagonal entries directly give the value of the corotating-frame frequency shift in units of $\omega_\text{dyn} = \sqrt{GM_*/R_*^3}$. Bottom row: the eigenvectors of (\ref{eq:om1_oblique}), i.e.~the coefficients of expansion of corotating-frame modes with respect to the basis of inertial-frame modes, in the case of alignment (left) and misalignment by an angle of $\pi/3$ (right). These values were calculated for a p-dominated mode of Model C with $n = -16$, $\ell = 1$. Rotation and field strength were set to their default values listed in Table \ref{tab:models}.}
  \label{fig:RBbox_l1}
\end{figure}

\begin{figure}
  \centering
  \includegraphics[width=\columnwidth]{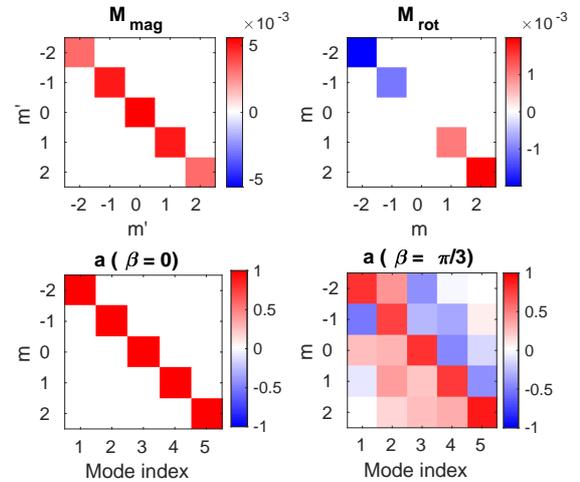}
  \caption{As for Fig.~\ref{fig:RBbox_l1}, but a mode with $n = -28$, $\ell = 2$.}
  \label{fig:RBbox_l2}
\end{figure}

\begin{figure}
  \centering
  \includegraphics[width=\columnwidth]{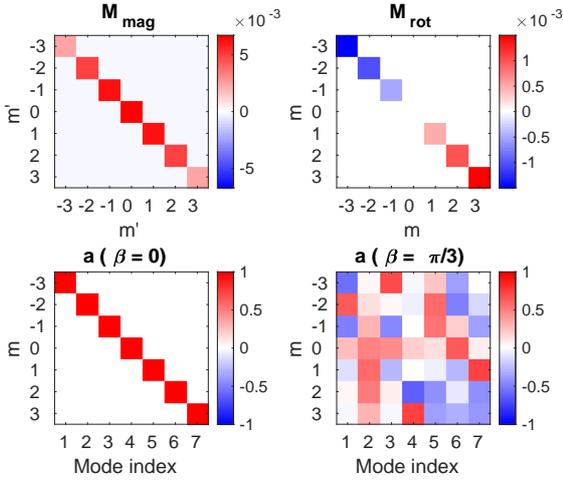}
  \caption{As for Fig.~\ref{fig:RBbox_l1}, but a mode with $n = -49$, $\ell = 3$.}
  \label{fig:RBbox_l3}
\end{figure}

The $m$-mixing process is illustrated in Figs~\ref{fig:RBbox_l1}--\ref{fig:RBbox_l3}, for a chosen mode of Model C with $\ell$ = 1, 2 and 3. The values of $\Omega$, $v_\text{A,cen}$ and $R_\text{f}$ are the same in all cases. Along the top row of each plot are the matrices $\boldsymbol{\mathcal{M}}_\text{mag}$ and $\boldsymbol{\mathcal{M}}_\text{rot}$, whose entries quantitatively represent the corotating-frame frequency shift that would be induced by each effect in the absence of the other. Along the bottom row are matrices containing the $2\ell + 1$ eigenvectors of (\ref{eq:om1_oblique}), where each column corresponds to one eigenvector. These are shown for two values of $\beta$. When $\beta = 0$ (aligned case) these yield the identity matrix (no $m$-mixing), but for $\beta = \pi/3$ (oblique case) the off-diagonal components are non-zero. Comparing the three values of $\ell$, it can be seen that the magnetic contribution to the frequency shifts, given by the components of $\boldsymbol{\mathcal{M}}_\text{mag}$, increases for higher $\ell$. Physically this can be understood from the fact that for modes possessing g-like character, larger $\ell$ are associated with smaller spatial scales and therefore larger Alfv\'{e}n frequencies, for the same $\omega_0$. In contrast, the rotational contribution, given by the components of $\boldsymbol{\mathcal{M}}_\text{rot}$, decreases for increasing $\ell$ given the same $m$. Consequently the amount of $m$-mixing increases with $\ell$, for the same obliquity.

\begin{figure*}
  \centering
  \includegraphics[width=\textwidth]{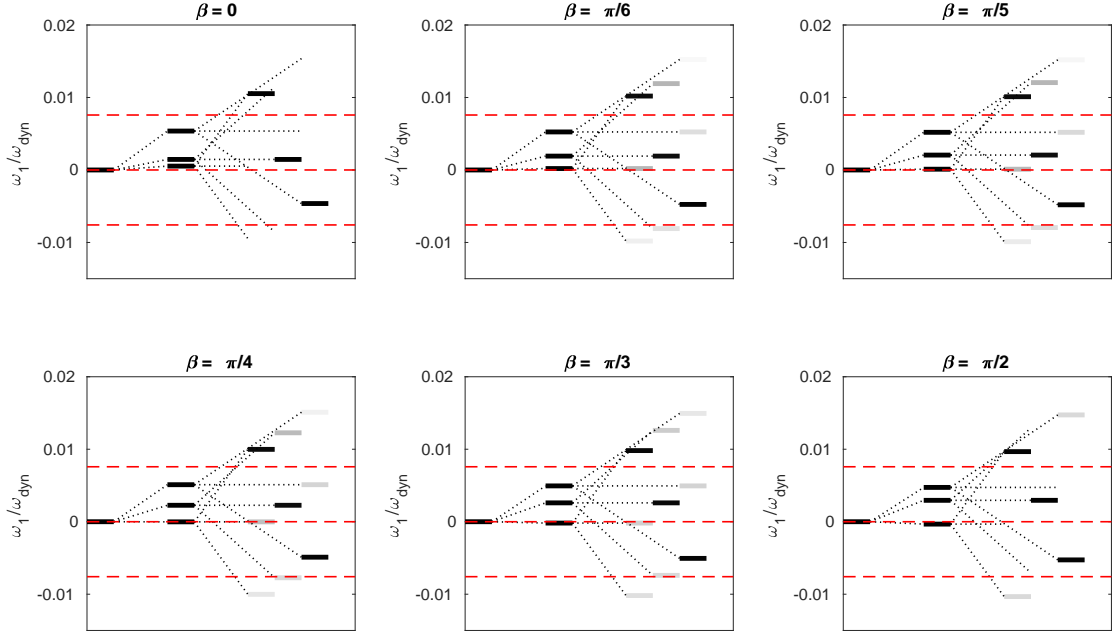}
  \caption{Splitting diagram for a p-dominated mode of Model C, with $n = -16$ and $\ell = 1$ (i.e.~that used in Fig.~\ref{fig:RBbox_l1}), under the default rotation and field strength listed in Table \ref{tab:models}, for six different values of the obliquity angle. Short, thick, horizontal black lines joined by thin dotted lines illustrate the splitting process, from the unperturbed mode (leftmost line at $\omega_1 = 0$) into a multiplet of $2\ell+1$ modes in the corotating frame (middle group), and then further into $(2\ell+1)^2$ modes in the inertial frame (rightmost group). In the rightmost group, the various modes have been horizontally offset for clarity. The different greyscale shades correspond to the absolute value of the associated $a_m$ coefficient, with black being 1 (maximum amplitude) and white being 0. For comparison, horizontal coloured dashed lines mark the position of the non-magnetic rotationally split multiplet in the inertial frame.}
  \label{fig:splittingC_l1_weakB}
\end{figure*}

\begin{figure*}
  \centering
  \includegraphics[width=\textwidth]{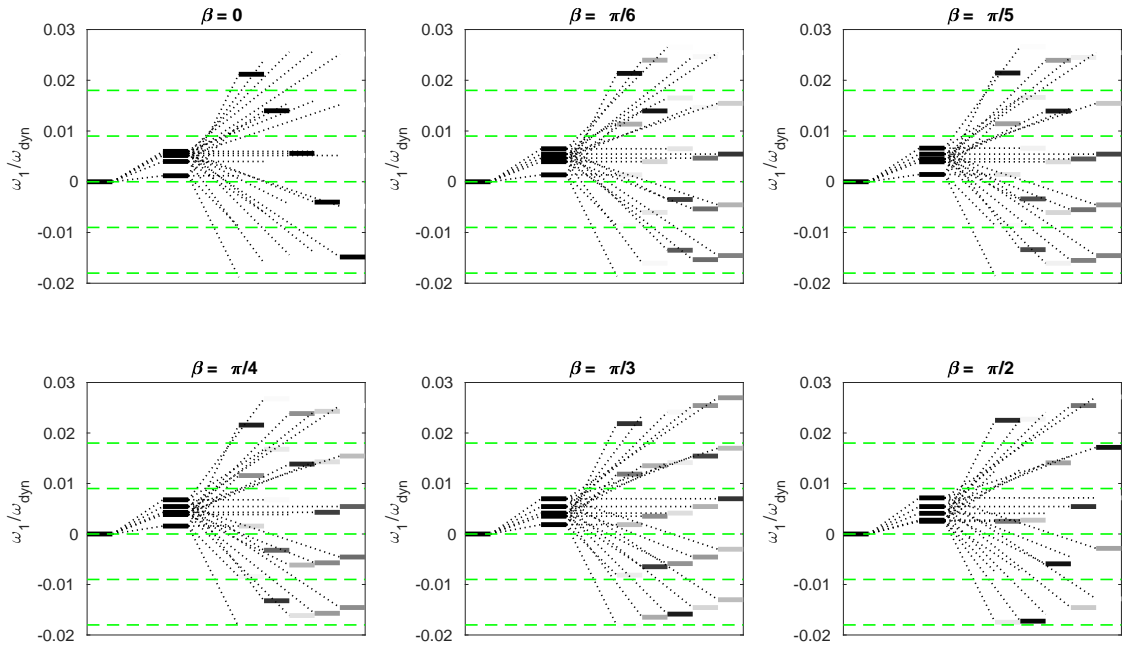}
  \caption{As for Fig.~\ref{fig:splittingC_l1_weakB}, but a mode with $n = -28$ and $\ell = 2$ (i.e.~that used in Fig.~\ref{fig:RBbox_l2}).}
  \label{fig:splittingC_l2_weakB}
\end{figure*}

\begin{figure*}
  \centering
  \includegraphics[width=\textwidth]{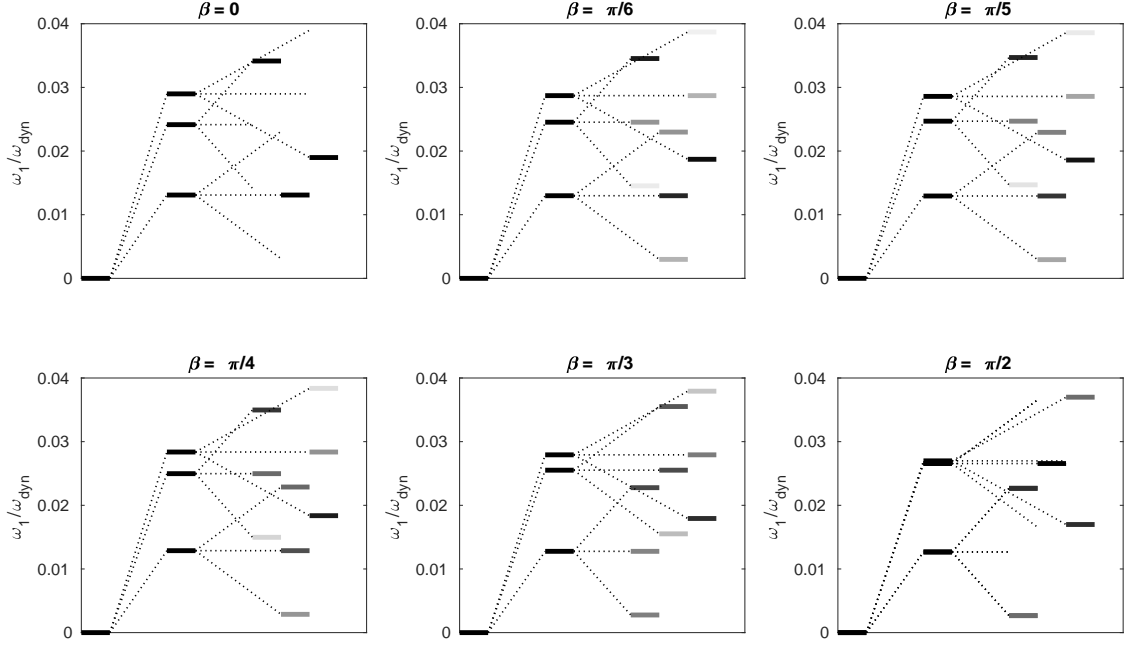}
  \caption{As for Fig.~\ref{fig:splittingC_l1_weakB}, but with a field strength three times higher.}
  \label{fig:splittingC_l1_strongB}
\end{figure*}

Figures \ref{fig:splittingC_l1_weakB}--\ref{fig:splittingC_l1_strongB} illustrate the splitting process for a chosen mode of Model C, for six different values of $\beta$ going from 0 (aligned) to $\pi/2$ (perpendicular). Similar plots for Models A and D can be found in Supplementary Figs S8--S13. It can be seen that greater $m$-mixing occurs for intermediate obliquities ($\beta \sim \pi/4$), higher $\ell$ and larger field strengths. In Fig.~\ref{fig:splittingC_l1_weakB}, it is apparent that in the inertial frame only $2\ell + 1$ out of $(2\ell + 1)^2$ peaks dominate, consistent with the $\mathbf{a}$ matrix remaining approximately diagonal even at substantial obliquities (see Fig.~\ref{fig:RBbox_l1}). Closer inspection of how the shape of this dominant sub-multiplet changes with obliquity reveals that while this may be asymmetric in general, it is still possible for this to appear nearly symmetric for some values of $\beta$ and have a spacing close to the value expected of pure rotation (Fig.~\ref{fig:splittingC_l1_weakB}, lower left panel). However, the centroid of this sub-multiplet is offset compared to the pure rotational multiplet. At larger field strengths (see Fig.~\ref{fig:splittingC_l1_strongB}), there may be many peaks of significant amplitude in the inertial frame. Notably, due to the heavy mixing, it would be possible to find symmetric sub-multiplets among these (e.g.~bottom row, middle panel).

\begin{figure*}
  \centering
  \includegraphics[width=0.8\textwidth]{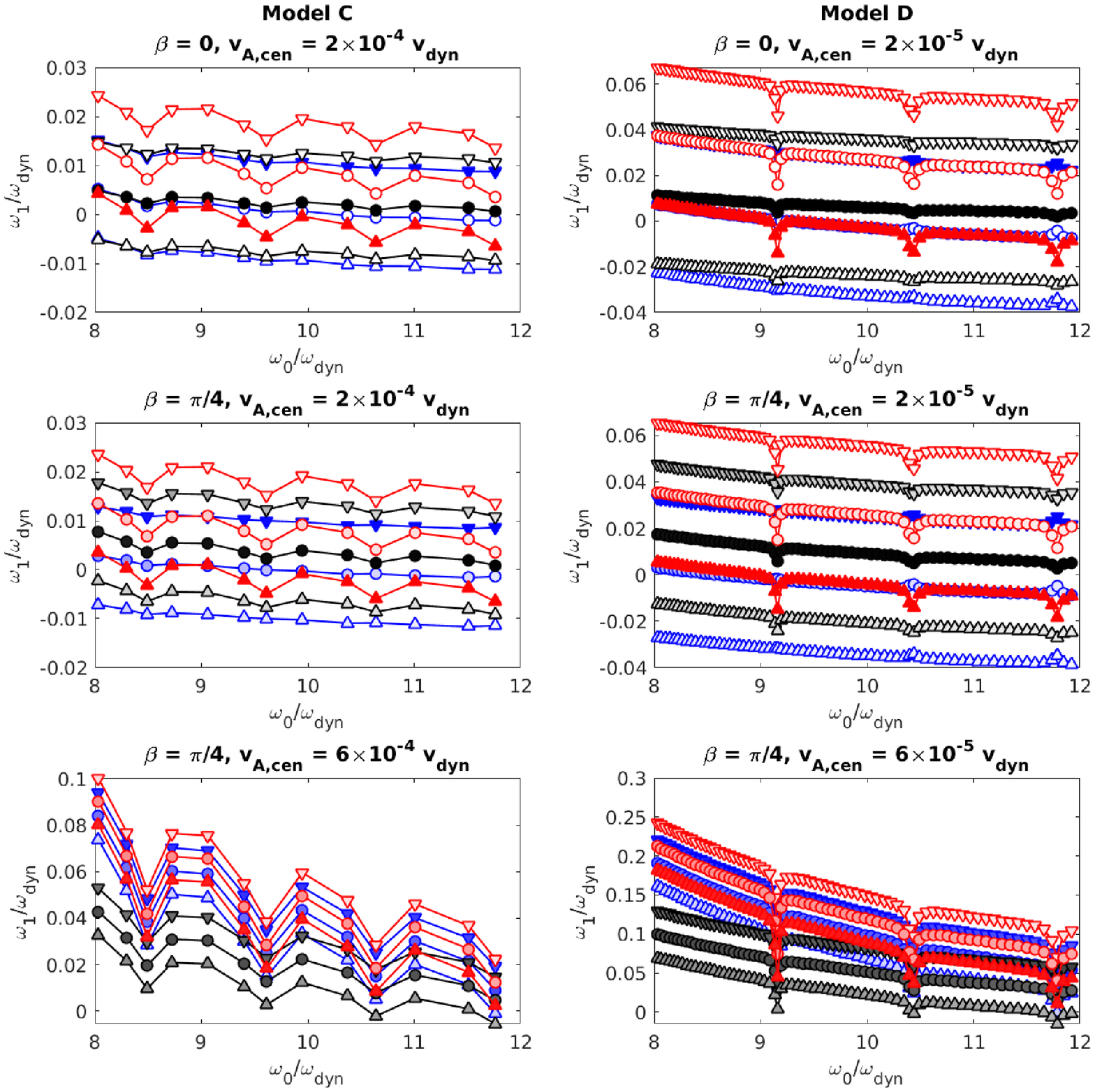}
  \caption{Inertial-frame frequency shifts versus unperturbed frequency, for all $\ell = 1$ modes of Models C (left) and D (right). Values of the field strength and obliquity are shown in the panel headers. Within each panel, the different colours correspond to the corotating-frame modes, which are each split further into $2\ell+1$ modes in the inertial frame. The different symbols represent the different $m$ components: upward triangles represent $m = +1$ (retrograde modes), circles represent $m = 0$ and downward triangles represent $m = -1$ (prograde modes). In addition, each symbol is filled with a colour whose saturation varies according to the value $|a_m|$. Fully saturated (i.e.~solid red/blue/black) corresponds to $|a_m| = 1$, while white corresponds to $|a_m| = 0$.}
  \label{fig:om1_vs_om0_CD_beta}
\end{figure*}

Frequency splittings for all $\ell = 1$ modes of Models C and D are summarised in Fig.~\ref{fig:om1_vs_om0_CD_beta}, where points are coloured according to the associated value of $|a_m|$. The increased $m$-mixing with larger obliquity and field strength can clearly be seen across all modes, both p- and g-dominated. The main difference between p- and g-dominated mixed modes appears to lie in the centroid frequency of the multiplet: this is located towards systematically higher values compared to the unperturbed value for g-dominated multiplets, a feature not present in the case of pure rotation (as seen in Figs~\ref{fig:comp_nofield_C_l12} and \ref{fig:comp_nofield_D_l1}), where the centroid offset is zero for all multiplets. This offset between p- and g-dominated multiplets increases towards lower frequencies, where magnetic effects become more significant.

\subsection{Radial extent of the field}\label{sec:results_Rf}
When constructing each Prendergast field solution, one has free choice over the parameter $R_\text{f}$, the radial extent of the field. The default values of $R_\text{f}$ for each model used in preceding sections are indicated in Table \ref{tab:models}, but in this particular section we choose to vary $R_\text{f}$ between 0.6--1.4 times the default, while keeping all other parameters (including $v_\text{A,cen}$) the same.

\begin{figure*}
  \centering
  \includegraphics[width=\textwidth]{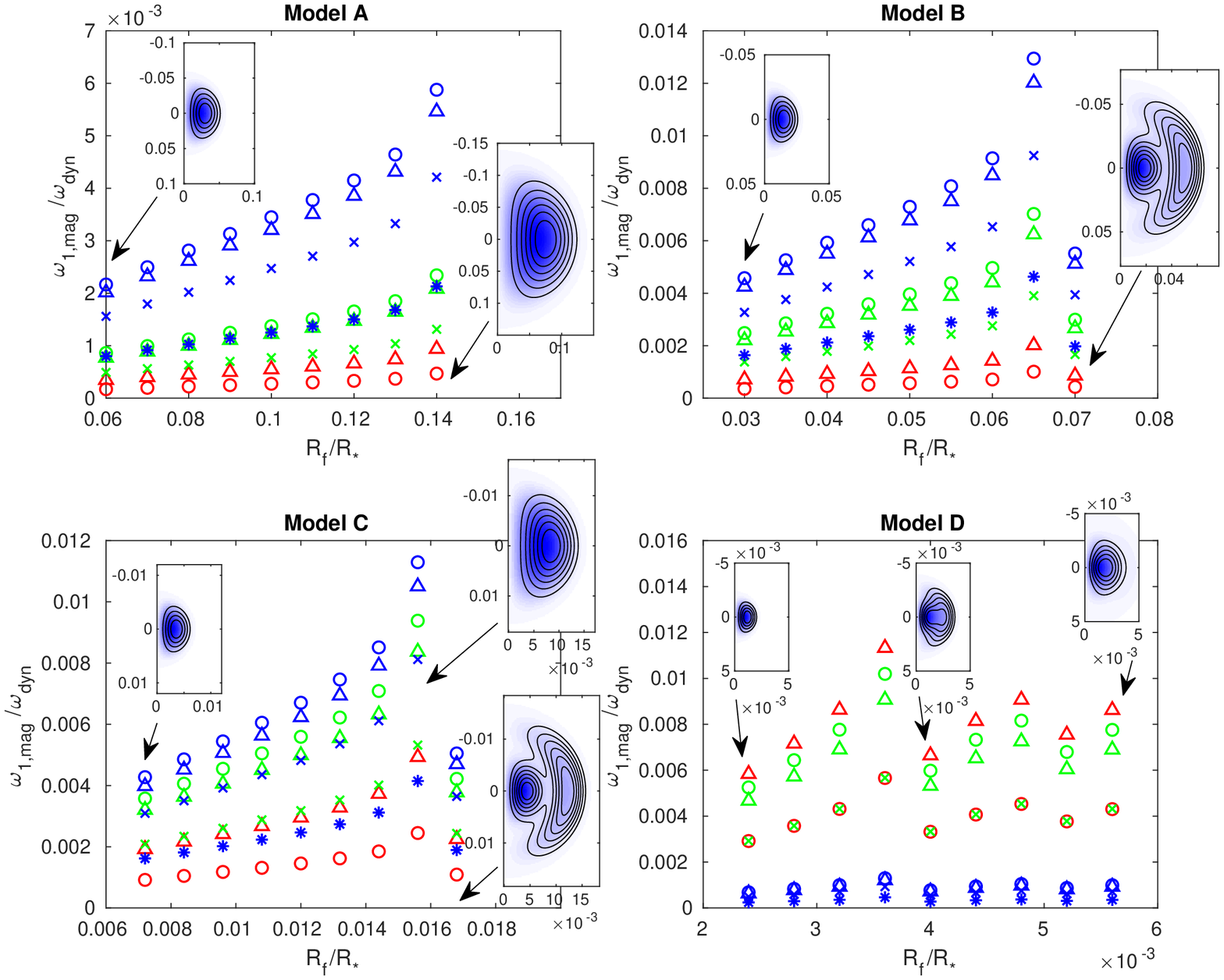}
  \caption{Magnetic contribution to the corotating-frame frequency shift (i.e.~the components of $\mathcal{M}_\text{mag}$) as a function of the radial extent $R_\text{f}$ of the field region, for selected modes of each model. Red, green and blue correspond to $\ell$ = 1, 2 and 3 respectively, while circles, triangles, crosses and asterisks correspond to $|m'|$ = 0, 1, 2 and 3 respectively. Inset plots show the field configurations for select values of $R_\text{f}$ indicated by the arrows. Each set of like-coloured points within each panel was generated for a single value of $n$. In order of increasing $\ell$, these were $n = -31, -50, -75$ for Model A, $n = -100, -179, -240$ for Model B, $n = -16, -28, -49$ for Model C and $n = -151, -255, -397$ for Model D. In the case of the two \textsc{mesa} models these choices of $n$ correspond to p-dominated modes (minima of Fig.~\ref{fig:inertias}).}
  \label{fig:Mmag_vs_Rfield}
\end{figure*}

Figure \ref{fig:Mmag_vs_Rfield} shows that the components of $\boldsymbol{\mathcal{M}}_\text{mag}$, representing the magnetic contribution to the overall frequency shift, increase in rough proportion with $R_\text{f}$. This may be explained by the integral in (\ref{eq:Lmag}) being larger when there is more volume of field to integrate over. However, this trend is not strict, and in some cases as for Model D the behaviour can be somewhat unpredicable. This appears to be tied to the complicating influence of variations in geometry/topology of the field as $R_\text{f}$ is modified; this occurs for the Prendergast model because of the dependence of (\ref{eq:lambda}) and (\ref{eq:Psi}) on $\rho$, whose shape over the interval $r \in [0, R_\text{f}]$ does not scale straightforwardly with $R_\text{f}$. At points where the configuration acquires additional radial structure (see inset panels), this tends to disrupt the trend. Further discussion of the field topology can be found in the next section.

This plot also sheds light on the effect of central condensation. Consider Models A and B, which are both polytropic models with the same $M_*$ and $R_*$ (and therefore dynamical frequency), but have different central condensations due to their differing polytropic indices. They also have similar central field strengths (differing by a factor of two, being larger in Model A). However, the magnetic shifts in Model B are larger than A by a factor of about two, and since this scales with $B^2$, if Model B had the same central field strength then the frequency shift would be $\sim$8 times larger than in Model A. The increased importance of magnetic effects at higher central condensations can be attributed to the smaller spatial scales of oscillation in the core due to the larger buoyancy frequency, leading to a higher Alfv\'{e}n frequency. This effect is more difficult to disentangle using Models C and D because the various modes differ wildly in their level of p-like versus g-like character, which represents a conflating influence. 

A minor further comment about Fig.~\ref{fig:Mmag_vs_Rfield} concerns the $\ell$-dependence for the different models. In Models A, B and C, for which the modes chosen for plotting all have relatively large amounts of mixed p- and g-like character, higher $\ell$ undergo larger shifts since the dominant factor is the shrinkage of spatial scales as $\ell$ increases (driving up the Alfv\'{e}n frequency). However, in Model D, which is more evolved than Model C, the much greater p-dominated character for higher $\ell$ means that the behaviour is reversed: higher $\ell$ values undergo smaller magnetic shifts.

\begin{figure}
  \centering
  \includegraphics[width=\columnwidth]{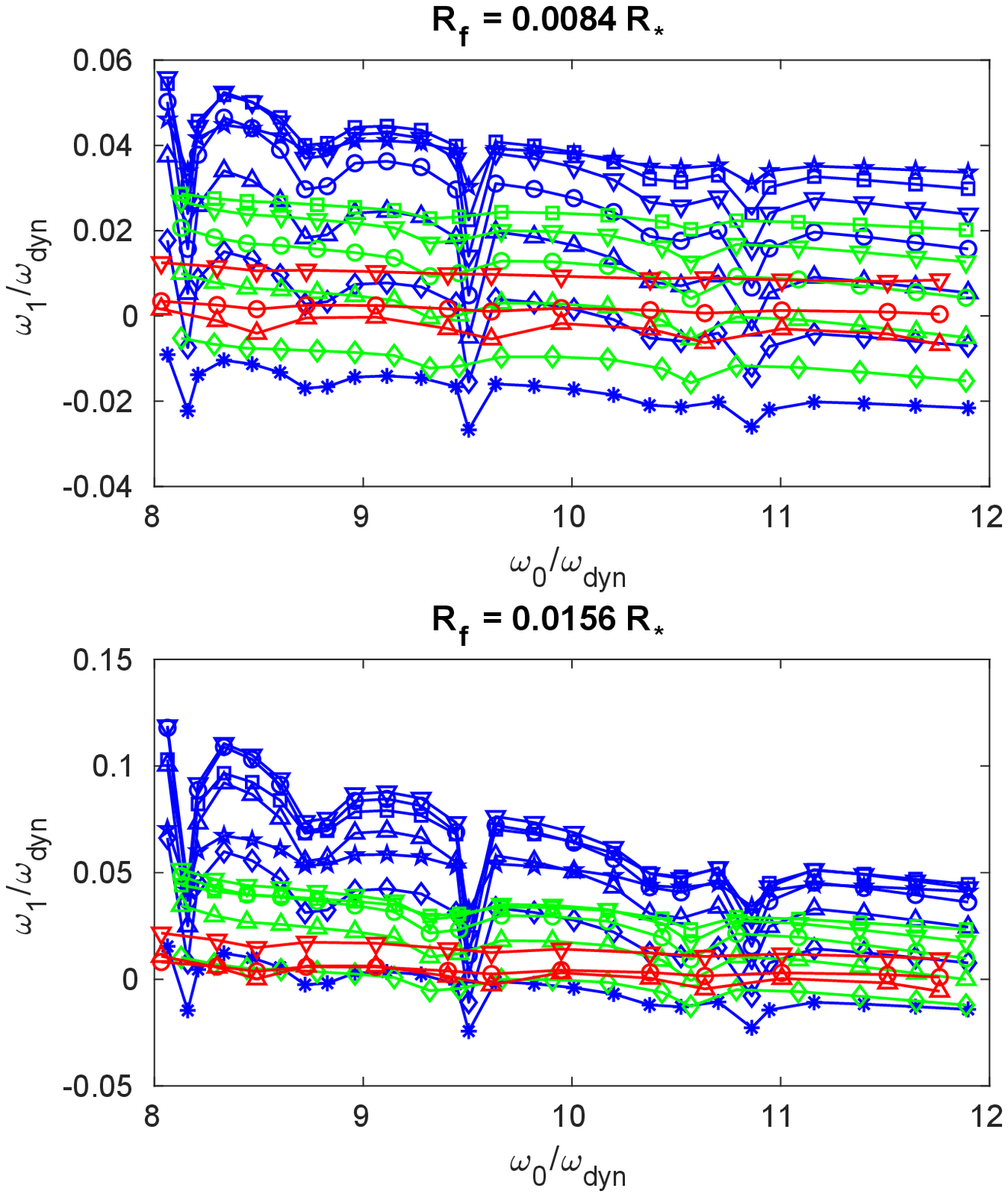}
  \caption{Inertial-frame frequency shifts versus unperturbed frequency for all modes of Model C, for two different values of $R_\text{f}$, these being 0.7 (top) and 1.3 (bottom) times the default value listed in Table \ref{tab:models}. Rotation and field strength are set to their default values, and $\beta = 0$. Red, green and blue correspond to $\ell$ = 1, 2 and 3 respectively, while different symbols correspond to different values of $m'$. In increasing order from $m' = -3$ to $+3$, these are denoted by pentagrams, squares, downward triangles, circles, upward triangles, diamonds, and asterisks respectively.}
  \label{fig:om1_vs_om0_C_Rfield}
\end{figure}

The inertial-frame frequency shifts (including the Coriolis force) for all modes of Model C are shown in Fig.~\ref{fig:om1_vs_om0_C_Rfield}, for two values of $R_\text{f}$. The g-dominated modes are seen to be much more sensitive to increases in $R_\text{f}$ compared to p-dominated modes, as can be seen by their larger values of $\omega_1$. This can be attributed to their preferential localisation to the core, where the field is located.

\subsection{Role of field topology}\label{sec:results_top}
\begin{figure*}
  \centering
  \includegraphics[width=\textwidth]{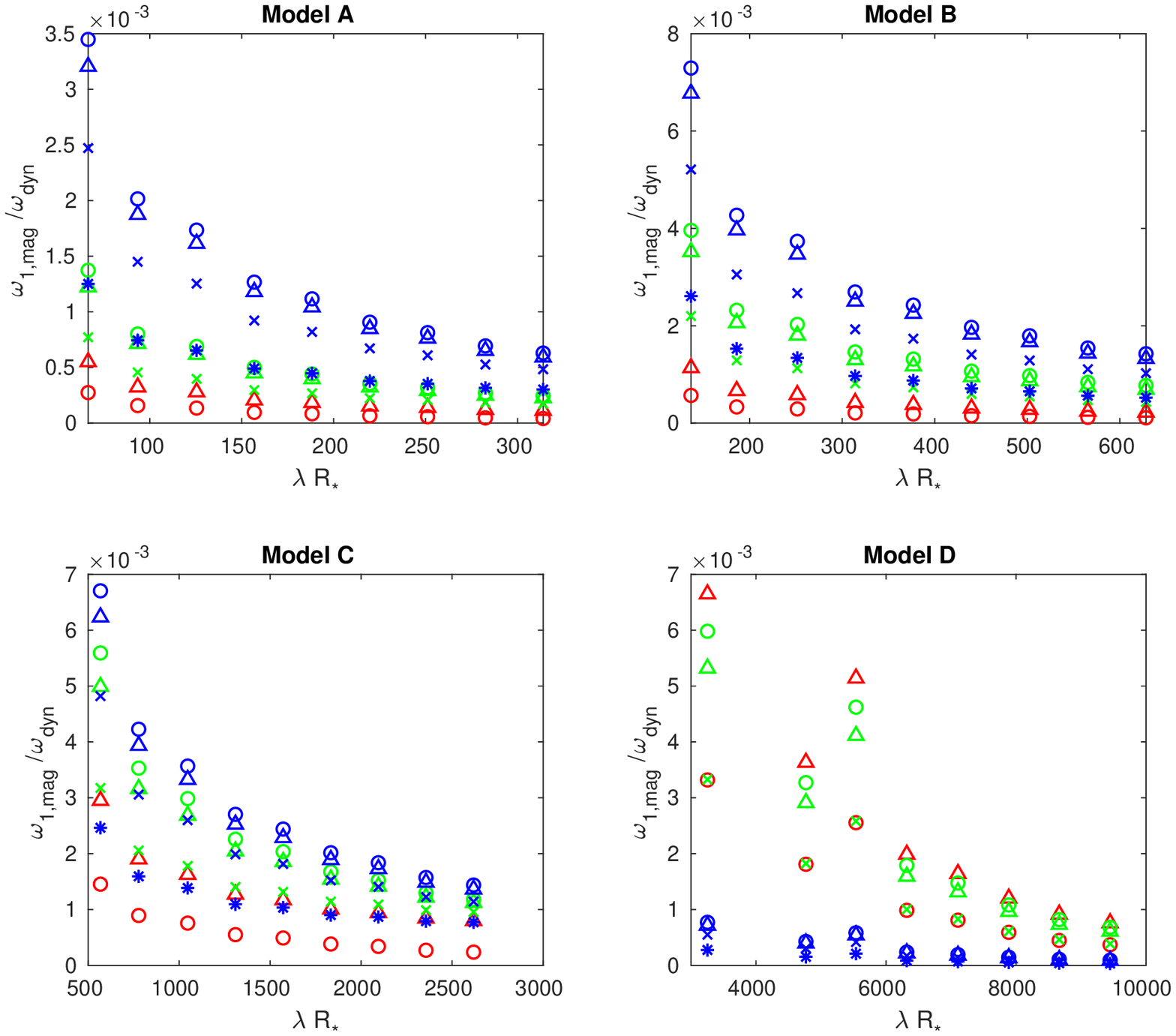}
  \caption{Magnetic contribution to the corotating-frame frequency shift (i.e.~the components of $\mathcal{M}_\text{mag}$) as a function of $\lambda$ expressed in units of $R_*^{-1}$. Larger values of $\lambda$ correspond to field configurations with more complex radial structure, shown in Fig.~\ref{fig:Prendergast_C} for Model C and in Supplementary Figs S1--S3 for the other models. Colours and marker styles are used in an identical manner to Fig.~\ref{fig:Mmag_vs_Rfield}, for the same selected modes.}
  \label{fig:Mmag_vs_lam}
\end{figure*}

As mentioned in Section \ref{sec:Prendergast}, multiple values of $\lambda$ can satisfy (\ref{eq:lambda}) for a given combination of $R_\text{f}$ and $\rho(r)$. These correspond to fields of different topologies (cf.~Fig.~\ref{fig:Prendergast_C}). Figure \ref{fig:Mmag_vs_lam} shows how the components of $\boldsymbol{\mathcal{M}}_\text{mag}$ vary as a function of $\lambda$. The general trend is for these to decrease with increasing $\lambda$, suggesting that fields with more radial structure may produce smaller magnetic shifts, even if the central field strength is held constant. For Model D the trend is not so smooth (second and fourth $\lambda$ values are outliers); inspection of Supplementary Figs S3 and S7 shows that these have unusual field structure, possessing a node/minimum very near the centre. The approach used here to control the field amplitude via scaling the central value thus produces anomalously large maximum field strengths and therefore frequency shifts. However, in all other cases the central and maximum field strengths are virtually identical. 

\begin{figure*}
  \centering
  \includegraphics[width=\textwidth]{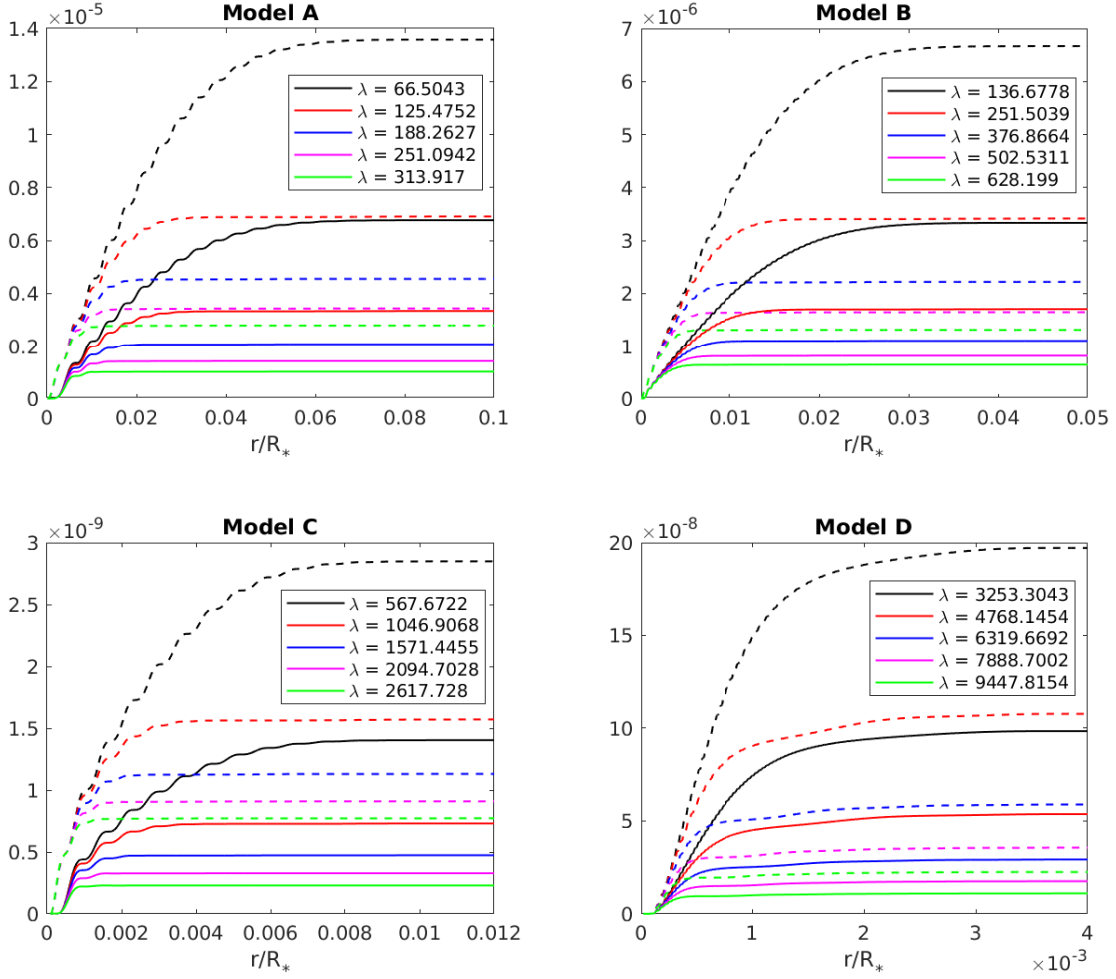}
  \caption{Values of the magnetic integral (\ref{eq:Lmag}) truncated at finite $r$, shown for selected $\ell = 1$ modes, with $n = -31$, $-100$, $-16$ and $-151$ for Models A, B, C and D, respectively. Different colours correspond to different values of $\lambda$ indicated in the legends. Solid lines correspond to $m' = 0$, while dashed lines correspond to $|m'| = 1$. The field radii $R_\text{f}$ are the same for each panel and correspond to the limits of the plot.}
  \label{fig:intmag_partial}
\end{figure*}

Further insight into the cause of this systematic behaviour with $\lambda$ comes from inspecting the spatial contribution of the integrand defined in (\ref{eq:Lmag}), by considering the value of the integral evaluated from 0 to some finite $r$ rather than all the way up to the stellar surface. Figure \ref{fig:intmag_partial} plots this as a function of the integration limit $r$, for the $\ell = 1$ modes used in Fig.~\ref{fig:Mmag_vs_lam}. While some small-scale fluctuations resulting from the mode structure are present, the overall behaviour in all cases is a relatively steep increase in the most central regions, where the rate of growth is largely independent of $\lambda$. Further out where the configurations undergo their finer-scale spatial reversals, the integrals plateau to a constant value that is lower for larger $\lambda$. Since all curves are monotonically increasing (if one ignores the small-scale fluctuations), this means that the integrand is always positive and so we conclude that for larger $\lambda$, it must be that the integrand itself is smaller over more of the volume. With aid of the plots showing total field strength (Figs S4--S7), it is apparent that this is driven by the steeper drop-off in field strength with radial distance for configurations with larger $\lambda$. Whether this is a peculiarity of the Prendergast solution or likely to be a property of real stellar magnetic fields is unknown. However, it is to be noted that a separate physical argument for weaker overall field strength in regions possessing smaller-scale field structure is the higher rate of Ohmic dissipation.

\begin{figure}
  \centering
  \includegraphics[width=\columnwidth]{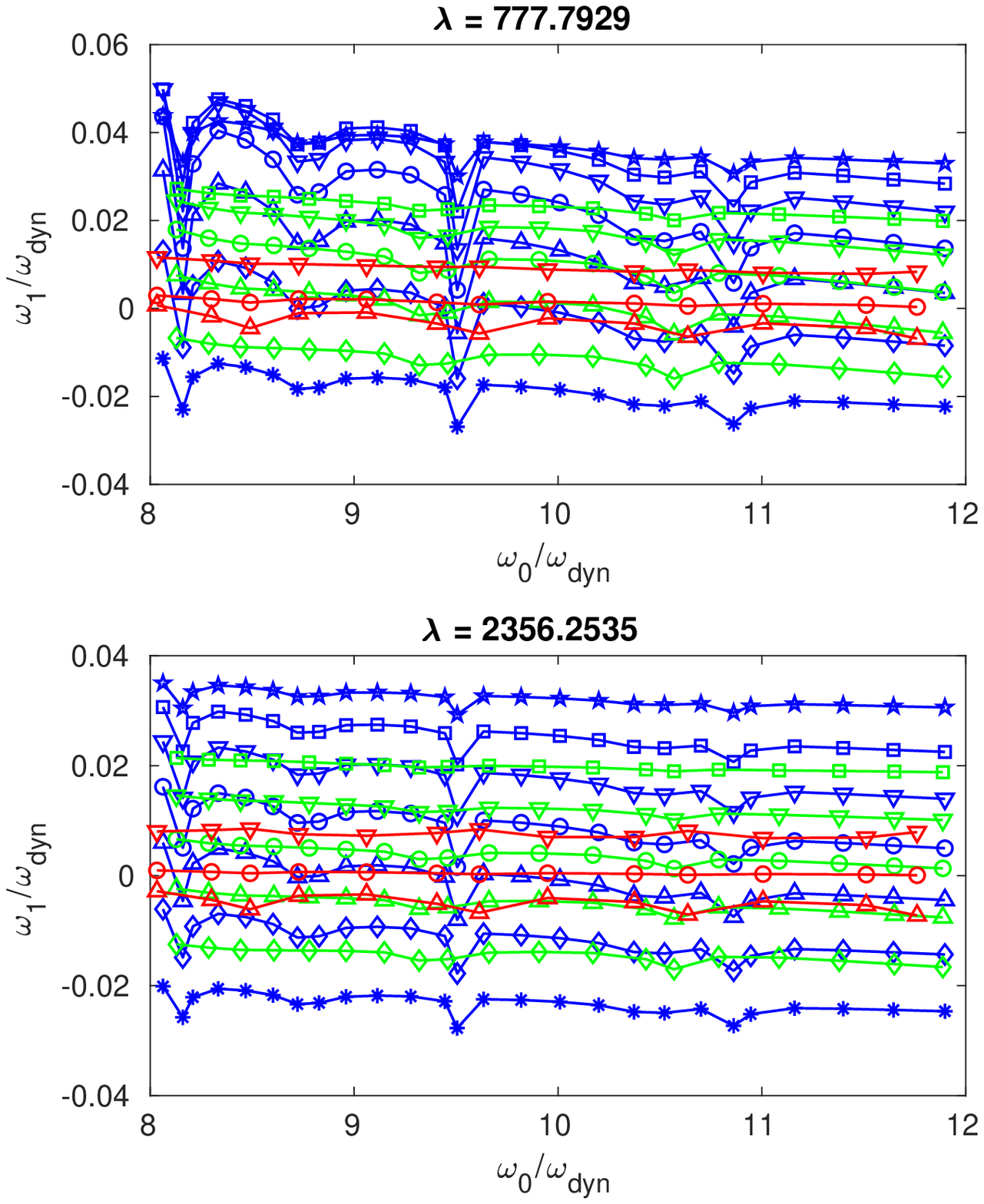}
  \caption{Inertial-frame frequency shifts versus unperturbed frequencies for all modes of Model C, for two different values of $\lambda$, these being the second (top) and eighth (bottom) roots of (\ref{eq:lambda}). Colours and marker styles are used in an identical manner to Fig.~\ref{fig:om1_vs_om0_C_Rfield}.}
  \label{fig:om1_vs_om0_C_lam}
\end{figure}

The inertial-frame frequency shifts (including the Coriolis force) for all modes of Model C are shown in Fig.~\ref{fig:om1_vs_om0_C_lam}, for two values of $\lambda$. As might be expected, similar to the case for $R_\text{f}$, the g-dominated mixed modes are far more sensitive to changes in $\lambda$ than the p-dominated ones.

\section{Discussion}\label{sec:discuss}
\subsection{Implications for asteroseismology}
Multiplet asymmetries are a known consequence of the Lorentz force, contrasting the Coriolis force which by itself only produces symmetric multiplets in the limit of slow rotation. While it may be tempting to use this signature as an indicator for the presence or absence of core magnetism, there are several complications. Firstly, magnetism is not the only possible cause of asymmetry \citep[cf.][]{Deheuvels2017}. Secondly, the results here show that it is possible for multiplets to appear symmetrically split, by a value close to that associated with pure rotation, even when the Lorentz force is comparable to the Coriolis force. This tends to occur for intermediate obliquities ($\beta \sim \pi/4$); see e.g.~Fig.~\ref{fig:splittingC_l1_weakB} and Fig.~\ref{fig:om1_vs_om0_CD_beta} (middle row). This result is important because it means that the detection of symmetric multiplets cannot alone be used to rule out the existence of a magnetic field.

A closely related concern is the fact that magnetic fields of non-zero obliquity are supposed to increase the multiplicity from $2\ell + 1$ to $(2\ell + 1)^2$ peaks in the inertial frame. Given this, it might be similarly tempting to interpret the observed lack of additional multiplicity as evidence of the absence of a magnetic field. However, it is clear from e.g.~Fig.~\ref{fig:RBbox_l1} (bottom right) that most of the additional components can have quite low amplitudes, perhaps an order of magnitude smaller than the dominant components, even in situations where the Lorentz force is comparable to the Coriolis force and $\beta$ is substantial. Thus in practice, many of the additional peaks may not stand out well above the noise, and only the $2\ell + 1$ dominant components along the diagonal of the $\mathbf{a}$ matrix have sufficient amplitudes to be identified. It is therefore possible that magnetic fields of substantial strengths are present in stars for which this scenario has previously been ruled out on the basis of the lack of multiplet asymmetry and/or absence of enhanced multiplicity. Rather, more detailed modelling on a case-by-case basis may be required to support such conclusions.

There may be some specific behaviours which could provide useful arguments for or against the existence of a core magnetic field. As noted throughout Section \ref{sec:results}, magnetic effects are more important for modes with shorter wavelengths and higher inertias. Signatures such as multiplet asymmetry and enhanced multiplicity are therefore expected to be most pronounced for modes that have lower $\omega_0$, higher $\ell$, are more g-dominated, and in stars with greater central condensations. In particular, the $1/\omega_0$ dependence of the magnetic shift and its tendency to systematically increase the centroid frequency of a multiplet is in contrast to the behaviour of rotation, which at leading order has no $\omega_0$ dependence and does not alter the centroid frequency. Since g-dominated mixed modes are more heavily affected by a core field, the centroids of their associated multiplets would be displaced further from the unperturbed values compared to p-dominated mixed modes. For example, if one were to perform forward modelling of an evolved star to reproduce well the p-dominated mode frequencies while neglecting the existence of a core field, then this would systematically underpredict the frequencies of g-dominated modes. The discrepancy would be worse at lower $\omega_0$ and higher $\ell$. Demonstration of such behaviour or lack thereof could provide a convincing case for/against core magnetism, but may be tricky in practice owing to the need to detect sufficient numbers of g-dominated mixed modes.

As a comment, this property of the Lorentz force to produce systematically positive frequency shifts owes to the fact that the integral in (\ref{eq:Mmag_el}) is positive definite as long as the integration limits extend beyond $R_\text{f}$ (outside of which $B = 0$). Upon integration by parts, the boundary terms may thence be neglected, and the integral can be approximated by $\int |\nabla \times (\bsxi \times \mathbf{B})|^2 \rmd V \geq 0$. Note that in this paper we choose to use the more general/exact expression in (\ref{eq:Mmag_el}): this is necessary in particular for Fig.~\ref{fig:intmag_partial}, since the radial integration limits are truncated within the field region.

Regarding seismic inference of the geometry/topology of a magnetic field, at first glance it may seem as though a near-infinite number of parameters would be needed to parametrise a field configuration of realistic complexity, making this a daunting prospect. However, this study shows how the problem may be made more tractable, since we have been able to parametrise a fairly complex magnetic field using just four global scalars: $\beta$, $v_\text{A,cen}$, $R_\text{f}$ and $\lambda$. Even without extensive characterisation of the g-dominated mixed modes, constraining some of these parameters may be possible using p-dominated mixed modes alone (cf.~Fig.~\ref{fig:Mmag_vs_Rfield}, for which the bottom row are p-dominated modes).

\subsection{Limitations}
The Prendergast solution may be a highly convenient modelling choice for the magnetic field, and supported by a large number of theoretical and empirical considerations, but ultimately it is difficult to know how accurately this would represent the field in the core of an actual star. As noted, for each stellar model there is actually a whole family of solutions, having different $\lambda$. There are suggestions that the configuration corresponding to the smallest $\lambda$ root is the one most likely to occur in reality, as it has the lowest associated energy \citep{Duez2010a}. However, it may also be argued that the topology of the final configuration should depend on that of the dynamo field at the point of cessation, assuming helicity conservation during relaxation as would be expected in ideal MHD. Thus it may be considerably more complex than the smallest-$\lambda$ configuration, although the substantial uncertainties associated with dynamo physics leave this issue open. In this work we have considered higher-$\lambda$ configurations and found these to produce smaller magnetic shifts, for the same central strength. This appears to be a consequence of a steeper rate of fall-off in field strength with radial distance, which is in turn linked to a larger fraction of the volume having small/near-zero field amplitude when there are more spatial reversals. While this might not seem inconceivable in reality, and would be favoured by physical arguments involving Ohmic dissipation anyhow, it is to be borne in mind that ultimately these results are for a specific type of field configuration, which may have certain peculiarities.

It is noticed that the rotational frequency shifts considered for Model D are comparable to the mode spacings, so it may be quasi-degenerate perturbation theory that would more appropriate. However, work done by \citet{Loi2019} suggests that first-order theory remains reasonably accurate at predicting the frequency shifts in this regime; rather, it is the accuracy of constructing the perturbations to the eigenfunctions (via the coefficients $c_k$) that suffers the most. However, that is not of interest here, and so we adopt first-order theory throughout. It is also worth commenting that the chosen field strength for Model D, which was adjusted so that magnetic and rotational frequency shifts would be comparable, is near the estimated critical strength at which gravity-Alfv\'{e}n wave resonance and mode conversion might occur. While the effects of this cannot be directly handled by perturbation theory, the work of \citet{Loi2020} found the existence of directions in which gravity wave propagation may not be significantly affected by the Lorentz force, and others where mode conversion (``trapping'') occurs efficiently. This anisotropy predicts that sectoral modes (with respect to the magnetic axis) may be relatively unaffected, while zonal modes are preferentially damped, which would act to depress certain $|a_m|$ coefficients, possibly pushing them below detection limits and thus lowering the apparent multiplicity. Further consideration of this possible ``hybrid'' approach for incorporating the effects of a dynamically significant field is beyond the scope of the current work, but may represent a possible avenue for future studies.

\section{Summary}\label{sec:summary}
In this paper we have presented a phenomenological study of the frequency splitting patterns induced by magnetism and rotation. We applied this to the case of evolved stars containing core magnetic fields, and examined how this is influenced by properties such as the obliquity and topology of the field. Realistic mixed configurations are used which are fully parametrisable by just four global scalars, demonstrating that seismic inference of the properties of a core magnetic field may be a reasonably tractable problem. It is found that greater radial complexity of the field (more fine-scale structure) leads to smaller magnetic contributions to the frequency shift, owing to a faster fall-off in field strength away from the centre. Topological variations in core fields among evolved stars, due to differing types of dynamos or otherwise, may therefore speculatively play a role in explaining some of the diversity in e.g.~mode depression behaviour.

A common assumption is that the lack of multiplet asymmetries and/or lack of increased multiplicity implies the absence of a magnetic field. However, the results of this work show that it is possible for both of these to be exhibited even when the Lorentz force is comparable to the Coriolis force, i.e.~magnetic and rotational contributions to the frequency splitting are comparable. This arises from the non-straightforward effects of obliquity and the nature of $m$-mixing, underscoring the subtlety of the Lorentz force and suggesting the need for more complete modelling of rotational and magnetic effects to support such conclusions.

Distinctive signatures of magnetic splitting that may allow it to be disentangled from rotational splitting include its general $1/\omega_0$ dependence, positive definite nature, and tendency to affect modes of small spatial scales the most (due to the $k$-dependence of the Alfv\'{e}n frequency, in contrast to rotational effects which are independent of $k$). Hence these signatures would be more pronounced for higher $\ell$, at greater central condensations, and among g-dominated mixed modes. It would be interesting to apply these concepts to analysing real objects in which previously, on the basis of fairly ``normal''-looking splitting patterns, magnetic fields may have been inferred not to exist. Note that the tools used in this study are publicly available (see Data Availability section below); those interested are encouraged to appropriate these for their own purposes.

\section*{Acknowledgements}
The author is supported by funding from Churchill College, Cambridge through a Junior Research Fellowship. We thank the \textsc{mesa} code development team members for all their efforts put into their publicly available stellar evolution code (version r11701).

\section*{Data availability}
Source codes and smaller data files can be found at the following link: \url{https://github.com/STCLoi/topology-obliquity-paper/}. For any queries, or access to larger data files, please contact the corresponding author.







\appendix

\section{\textsc{mesa} inlist}\label{sec:inlist}
Models C and D were generated by \textsc{mesa} (r11701) using the following inlist, and correspond to output profile numbers 26 and 31 thereof:
\begin{verbatim}
&star_job
  ! begin with a pre-main sequence model
    create_pre_main_sequence_model = .true.

  ! save a model at the end of the run
    save_model_when_terminate = .true.
    save_model_filename = `2Msun.mod'

  ! display on-screen plots
    pgstar_flag = .true.

/ !end of star_job namelist

&controls
  ! starting specifications
    initial_mass = 2 ! in Msun units

  ! stopping condition
    max_age = 1.01d9

  ! mesh adjustment
    mesh_delta_coeff = 0.2d0
    max_dq = 1d-3

/ ! end of controls namelist
\end{verbatim}

\section{Perturbed current density}\label{sec:Jcompts}
Below are the components of $\mathbf{J}'(r, \theta', \phi')$, the Eulerian perturbation to the current density, which appears in the expression (\ref{eq:Mmag_el}) for the components of the matrix $\boldsymbol{\mathcal{M}}_\text{mag}$. These are for the special case of the Prendergast solution (see Section \ref{sec:Prendergast} for details). Arguments have been suppressed for brevity, but recall that $R_\ell$, $H_\ell$ and $\Psi$ are sole functions of $r$, and $Y_\ell^{m'}$ is a function of $\theta'$ and $\phi'$.
\begin{align}
  J_r' &= \frac{1}{r} \left\{ A_{\phi'} \cot \theta' - \rmi m' \frac{A_{\theta'}}{\sin \theta'} + \left[ \frac{2\rmi m'}{r^2 \sin^2 \theta'} \left( \frac{H_\ell \Psi}{r} - H_\ell \frac{\rmd \Psi}{\rmd r} \right. \right. \right. \nonumber \\
    & \left. - \Psi \frac{\rmd H_\ell}{\rmd r} \right) + \frac{\lambda \Psi \cos \theta'}{r} \left( \frac{\rmd R_\ell}{\rmd r} - \frac{H_\ell}{r} \left( \frac{m'^2}{\sin^2 \theta'} + \ell(\ell+1) \right) \right) \nonumber \\
    & \left. + \frac{\lambda R_\ell \cos \theta'}{r} \frac{\rmd \Psi}{\rmd r} \right] Y_\ell^{m'} + \left[ 2\rmi m' \frac{\cot \theta'}{r^2} \left( \Psi \frac{\rmd H_\ell}{\rmd r} - \frac{\Psi H_\ell}{r} \right. \right. \nonumber \\
    & \left. + H_\ell \frac{\rmd \Psi}{\rmd r} \right) - \frac{\lambda \Psi H_\ell}{r^2} \left( \ell(\ell+1) \sin \theta' - \frac{m'^2}{\sin \theta'} \right) \nonumber \\
    & \left. \left. + \frac{\lambda \sin \theta'}{r} \left( \Psi \frac{\rmd R_\ell}{\rmd r} + R_\ell \frac{\rmd \Psi}{\rmd r} \right) \right] \frac{\del Y_\ell^{m'}}{\del \theta'} - \frac{\rmi m' H_\ell}{r^2} \frac{\rmd \Psi}{\rmd r} \frac{\del^2 Y_\ell^{m'}}{\del \theta'^2} \right\} \label{eq:Jr}
\end{align}
\begin{align}
  J_{\theta'}' &= \frac{\rmi m'}{r \sin \theta'} A_r - \left[ \frac{2\rmi m' \cot \theta'}{r^2} \left( 2 \left( \frac{\rmd H_\ell}{\rmd r} - \frac{H_\ell}{r} \right) \left( \frac{\rmd \Psi}{\rmd r} - \frac{\Psi}{r} \right) \right. \right. \nonumber \\
    & \left. + \Psi \frac{\rmd^2 H_\ell}{\rmd r^2} + H_\ell \frac{\rmd^2 \Psi}{\rmd r^2} \right) - \frac{\lambda \sin \theta'}{r^2} \left( H_\ell \frac{\rmd \Psi}{\rmd r} - \frac{H_\ell \Psi}{r} + \Psi \frac{\rmd H_\ell}{\rmd r} \right) \nonumber \\
    & + \left( \ell(\ell+1) - \frac{m'^2}{\sin^2 \theta'} \right) + \frac{\lambda \sin \theta'}{r} \left( \Psi \frac{\rmd^2 R_\ell}{\rmd r^2} + R_\ell \frac{\rmd^2 \Psi}{\rmd r^2} \right. \nonumber \\
    & \left. \left. + 2 \frac{\rmd \Psi}{\rmd r} \frac{\rmd R_\ell}{\rmd r} \right) \right] Y_\ell^{m'} + \frac{\rmi m'}{r^2} \left( \frac{\rmd^2 \Psi}{\rmd r^2} H_\ell - \frac{\rmd \Psi}{\rmd r} \frac{H_\ell}{r} + \frac{\rmd \Psi}{\rmd r} \frac{\rmd H_\ell}{\rmd r} \right) \frac{\del Y_\ell^{m'}}{\del \theta'} \label{eq:Jth}
\end{align}
\begin{align}
  J_{\phi'}' &= \frac{\sin \theta'}{r} \left[ \frac{\rmd \Psi}{\rmd r} \frac{\rmd^2 R_\ell}{\rmd r^2} + 2\frac{\rmd^2 \Psi}{\rmd r^2} \frac{\rmd R_\ell}{\rmd r} - \frac{2R_\ell}{r^2} \frac{\rmd \Psi}{\rmd r} + R_\ell \frac{\rmd^3 \Psi}{\rmd r^3} \right. \nonumber \\
    & \left. - \frac{\ell(\ell+1)}{r} \left( H_\ell \frac{\rmd^2 \Psi}{\rmd r^2} - \frac{H_\ell}{r} \frac{\rmd \Psi}{\rmd r} + \frac{\rmd H_\ell}{\rmd r} \frac{\rmd \Psi}{\rmd r} - \frac{2 \Psi H_\ell}{r^2} \right) \right] Y_\ell^{m'} \nonumber \\
    & + \left[ \frac{\cos \theta'}{r^2} \left( 3 \frac{\rmd \Psi}{\rmd r} \frac{\rmd H_\ell}{\rmd r} + \frac{3(R_\ell - H_\ell)}{r} \frac{\rmd \Psi}{\rmd r} - \frac{4\Psi}{r} \frac{\rmd H_\ell}{\rmd r} \right. \right. \nonumber \\
    & \left. + 2(1 - \ell(\ell+1)) \frac{\Psi H_\ell}{r^2} + 2\Psi \frac{\rmd^2 H_\ell}{\rmd r^2} + H_\ell \frac{\rmd^2 \Psi}{\rmd r^2} \right) \nonumber \\
    & \left. - \frac{\rmi m' \lambda}{r^2} \left( H_\ell \left( \frac{\rmd \Psi}{\rmd r} - \frac{\Psi}{r} \right) + \Psi \left( \frac{\rmd H_\ell}{\rmd r} - \frac{R_\ell}{r} \right) \right) \right] \frac{\del Y_\ell^{m'}}{\del \theta'} \nonumber \\
    & - \frac{\sin \theta'}{r^2} \left[ H_\ell \left( \frac{\rmd^2 \Psi}{\rmd r^2} - \frac{1}{r} \frac{\rmd \Psi}{\rmd r} + \frac{2\Psi}{r^2} \right) + \frac{\rmd \Psi}{\rmd r} \left( \frac{\rmd H_\ell}{\rmd r} - \frac{R_\ell}{r} \right) \right] \frac{\del^2 Y_\ell^{m'}}{\del \theta'^2} \label{eq:Jph}
\end{align}
where
\begin{align}
  A_r &= -\frac{1}{r^2} \left[ \rmi m' \lambda \Psi R_\ell + 2 \cos \theta' \left( R_\ell \frac{\rmd \Psi}{\rmd r} - \frac{\ell(\ell+1)}{r} H_\ell \Psi \right) \right] Y_\ell^{m'} \nonumber \\
  & + \frac{\sin \theta'}{r^2} \left[ \frac{2\Psi H_\ell}{r} - \frac{\rmd \Psi}{\rmd r} R_\ell \right] \frac{\del Y_\ell^{m'}}{\del \theta'} \label{eq:Ar}
\end{align}
\begin{align}
  A_{\theta'} &= \frac{\sin \theta'}{r} \left[ \frac{\rmd \Psi}{\rmd r} \frac{\rmd R_\ell}{\rmd r} - \frac{\ell(\ell+1)}{r} \frac{\rmd \Psi}{\rmd r} H_\ell + R_\ell \frac{\rmd^2 \Psi}{\rmd r^2} \right] Y_\ell^{m'} \nonumber \\
  & + \frac{1}{r^2} \left[ \cos \theta' \left( 2\Psi \frac{\rmd H_\ell}{\rmd r} + H_\ell \frac{\rmd \Psi}{\rmd r} - \frac{2H_\ell \Psi}{r} \right) \right. \nonumber \\
    & \left. - \rmi m' \lambda \Psi H_\ell \right] \frac{\del Y_\ell^{m'}}{\del \theta'} - \frac{H_\ell \sin \theta'}{r^2} \frac{\rmd \Psi}{\rmd r} \frac{\del^2 Y_\ell^{m'}}{\del \theta'^2} \label{eq:Ath}
\end{align}
\begin{align}
  A_{\phi'} &= \frac{1}{r} \left[ 2\rmi m' \frac{\cot \theta'}{r} \left( \frac{\rmd H_\ell}{\rmd r} \Psi + \frac{\rmd \Psi}{\rmd r} H_\ell - \frac{H_\ell \Psi}{r} \right) \right. \nonumber \\
    & + m'^2 \lambda \frac{\Psi H_\ell}{r \sin\theta'} + \lambda \sin \theta' \left( \Psi \frac{\rmd R_\ell}{\rmd r} \right. \nonumber \\
    & \left. \left. - \ell(\ell+1) \frac{\Psi H_\ell}{r} + R_\ell \frac{\rmd \Psi}{\rmd r} \right) \right] Y_\ell^{m'} - \frac{\rmi m' H_\ell}{r^2} \frac{\rmd \Psi}{\rmd r} \frac{\del Y_\ell^{m'}}{\del \theta'} \label{eq:Aph}
\end{align}


\bsp	
\label{lastpage}
\end{document}